\documentclass[11pt,letterpaper]{article}

\usepackage[letterpaper,margin=1in]{geometry}
\usepackage{amsmath,amssymb,amsthm,mathtools}
\usepackage{thmtools,thm-restate}
\usepackage{graphicx}
\usepackage{booktabs}
\usepackage{enumitem}
\usepackage{xcolor}
\usepackage{microtype}
\usepackage[hidelinks]{hyperref}
\usepackage[nameinlink,capitalize,noabbrev]{cleveref}
\Crefname{appendix}{Appendix}{Appendices}
\usepackage[backend=biber,style=numeric,maxbibnames=99]{biblatex}
\usepackage{bbm}
\usepackage{todonotes}
\usepackage[normalem]{ulem}

\setlist{leftmargin=*}
\numberwithin{equation}{section}

\theoremstyle{plain}
\newtheorem{theorem}{Theorem}[section]

\newtheorem{lemma}[theorem]{Lemma}
\newtheorem{corollary}[theorem]{Corollary}
\newtheorem{proposition}[theorem]{Proposition}
\newtheorem{claim}[theorem]{Claim}
\newtheorem{observation}[theorem]{Observation}
\crefname{observation}{Observation}{Observations}
\Crefname{observation}{Observation}{Observations}

\theoremstyle{definition}
\newtheorem{definition}[theorem]{Definition}

\theoremstyle{remark}

\newcommand{\defn}[1]{\textbf{\emph{#1}}}
\renewcommand{\paragraph}[1]{\vspace{.2cm}\noindent\textbf{#1}}

\DeclareMathOperator{\load}{load}

\DeclareMathOperator{\Vol}{Vol}
\DeclareMathOperator{\Enc}{Enc}

\newcommand{\eps}{\varepsilon}
\renewcommand{\epsilon}{\varepsilon}
\newcommand{\R}{\mathbb{R}}
\newcommand{\N}{\mathbb{N}}
\newcommand{\Z}{\mathbb{Z}}
\newcommand{\E}{\mathbb{E}}

\newcommand{\ind}{\mathbf{1}}

\DeclareRobustCommand{\copen}{\textnormal{\textsc{open}}}
\DeclareRobustCommand{\cclose}{\textnormal{\textsc{close}}}
\DeclareRobustCommand{\copenclose}{\textnormal{\textsc{open-close}}}
\DeclareRobustCommand{\cmiddle}{\textnormal{\textsc{middle}}}
\DeclareRobustCommand{\cempty}{\textnormal{\textsc{empty}}}

\newcommand{\reinflate}{\text{\textsc{Reinflate}}}
\newcommand{\query}{\text{\textsc{Query}}}
\newcommand{\ins}{\text{\textsc{Insert}}}
\newcommand{\delete}{\text{\textsc{Delete}}}

\newcommand{\HT}{\mathcal{H}}
\newcommand{\CHT}{\mathcal{C}}

\newcommand{\csmall}{c_{\mathrm{small}}}
\newcommand{\backup}{\mathcal{B}}
\newcommand{\member}{\mathcal{M}}
\newcommand{\ID}{\textrm{ID}}
\newcommand{\DHT}{\mathcal{D}}

\newcommand{\pifsch}{\pi^\textrm{FSCH}}

\definecolor{alexpurple}{HTML}{800080}
\definecolor{hannapink}{HTML}{DA3287}
\definecolor{billblue}{HTML}{0000FF}
\newif\ifdraftcomments
\draftcommentstrue

\title{Catalan Hashing: \\ Efficient Linear Probing for Variable-Length Items}
\author{\large Alex Conway \\ \normalsize Cornell University \\ \small\texttt{me@ajhconway.com}
  \and
\large Hanna Koml\'os \\ \normalsize Max Planck Institute \\ \normalsize for Informatics \\ \small\texttt{hkomlos@gmail.com}
\and
\large William Kuszmaul \\ \normalsize Carnegie Mellon \\ \normalsize University \\ \small\texttt{kuszmaul@cmu.edu}
}

\date{}

\begin{document}

\hypersetup{pageanchor=false}
\begin{titlepage}
\thispagestyle{empty}
\maketitle

\begin{abstract}
A defining feature of the linear-probing hash table is its spatial locality: at load factor $1 - \varepsilon$, the hash table can be implemented so that each operation scans a single contiguous region of expected size $O(1/\varepsilon)$ (this is known as the \emph{scan cost}). This guarantee, however, relies on all items having unit size. In this paper, we ask whether it is possible to achieve a similar guarantee, but in a hash table that stores items of arbitrary lengths $|x| \ge 1$.

The main result of the paper is a new hash table called \emph{Catalan hashing}, which supports amortized expected scan cost $O(|x|/\varepsilon)$ for positive queries, $O(1/\varepsilon)$ for negative queries, and $O(|x|\varepsilon^{-1}\log^{1+\delta}|x|)$ for insertions/deletions, for any constant $\delta > 0$. As a corollary, we obtain the same bounds in the cache-oblivious model, with a scan cost of $S$ translating to $1 + O(S/B)$ I/Os. Along the way, we introduce several high-level techniques that may be of independent interest; this includes a novel way of using erasure codes in order to perform write optimization within a hash table, and a connection to scheduling that allows us to place elements in a way which avoids clustering at high load factors.
\end{abstract}

\end{titlepage}

\hypersetup{pageanchor=true}
\setcounter{page}{1}

\section{Introduction}
\label{sec:intro}

First introduced in the 1950s~\cite{peterson57,Knuth63note}, the linear-probing hash table has stood for more than half a century as one of the simplest and most widely used hash table designs. In its most basic form, the hash table works as follows. Elements live in an array of some size $m$, and each element $x$ hashes to a uniformly random position $h(x)$ in that array. To insert an item $x$, we examine cells $h(x), h(x) + 1, h(x) + 2,\ldots \pmod m$ until we find a free cell where we place the item; and, likewise, to query an item $x$, we scan the cells $h(x), h(x) +1, h(x) + 2, \ldots \pmod m$ and until we either find $x$, or until we encounter a free cell (meaning that $x$ is not present).

What makes linear probing special---both in practice and in theory---is its \emph{spatial locality}~\cite{Knuth98,PaghPR09,BenderKuszmaulKuszmaul2021GraveyardHashing,BravermanKuszmaul2025LinearProbing,FarachColtonKrapivinKuszmaul2024OpenAddressing,RichterAD15}: each operation accesses just one small contiguous region of the hash table. We refer to the size of this region as the \emph{scan cost} of the operation. (This quantity has also been called the \emph{locality} of the operation~\cite{Zamir2026Locality}).

At constant load factor $1 - \Theta(1)$, the expected scan cost per operation is $O(1)$. When the hash table is filled to a high load factor $1 - \eps$ for a small parameter $\eps \in (0, 1)$, naive linear probing suffers from a phenomenon known as primary clustering~\cite{Knuth63note,KonheimWeiss1966Occupancy,Knuth98}, which increases the worst-case expected scan cost to $\Theta(\eps^{-2})$. Fortunately, this clustering effect can actually be eliminated with the help of additional algorithmic optimizations (the use of the Robin-Hood ordering to improve queries \cite{AmbleKnuth74,CelisLarsonMunro1985RobinHood} and the use of graveyard hashing~\cite{BenderKuszmaulKuszmaul2021GraveyardHashing} to improve insertions). With these optimizations in place, the expected scan cost per operation is $O(\eps^{-1})$.

The above results all operate under the assumption that items have a fixed size. In many applications, however, this is not the case: An item $x$ may occupy an arbitrary number $|x| \ge 1$ of words (representing, for instance, a variable-length key together with a variable-length value). A common workaround is to use \emph{indirection}: store a fixed-size pointer in the hash table and allocate the item itself elsewhere in memory. But indirection defeats the very spatial locality that makes linear probing so desirable---each operation must now access multiple separate regions of memory. 
%

This raises a natural question: Can linear probing be adapted to store variable-length items \emph{inline}, so that the expected scan cost per query is close to $O(\eps^{-1} \cdot |x|)$, where $|x|$ is the length of the item being queried? In other words, does there exist a data structure that behaves \emph{like} linear probing, but that supports items of arbitrary sizes?

Such a data structure would also be of direct interest in the \emph{external-memory} model~\cite{AggarwalVitter1988IO}, where memory is divided into blocks of size $B$ and the cost of an operation is the number of blocks it accesses (called \emph{I/Os}). In this model, a linear-probing operation with scan cost $S$ incurs $1 + O(S/B)$ I/Os in expectation, so bounds on scan cost translate directly into bounds on I/Os. Thus, another way to frame our question is simply: does there exist a cache-oblivious\footnote{A data structure is \emph{cache-oblivious} if it does not need to know the block size $B$ or cache size \(M\) in advance.} external-memory data structure in which querying an item $x$ costs $1 + O(\eps^{-1}|x| / B)$ I/Os in expectation?

\paragraph{This Paper: Catalan Hashing.}
In this paper, we introduce \emph{Catalan hashing}, which successfully extends linear probing to support variable-length items. In the final version of the data structure (\Cref{thm:intro-dynamic-high}), we are able to support queries with an amortized expected scan cost of 
$$O(\eps^{-1} \cdot |x|),$$
and insertions and deletions with an amortized expected scan cost of 
$$O(\eps^{-1} \cdot |x| \cdot \log^{1+\delta} |x|),$$
where $\delta > 0$ is a positive constant of our choice. The query bound applies to both positive queries (for elements that are present) and negative ones (for elements that are absent). In fact, for negative queries, we are able to achieve an even stronger expected scan-cost bound of $O(\eps^{-1})$, independent of the item size $|x|$.

We develop our results systematically in five parts.
Parts~1--3 cover static and then dynamic constructions in the constant-load-factor regime (load factor $1 - \Theta(1)$).
Parts~4--5 then extend these constructions to the high-load-factor regime (load factor $1 - \eps$). We state our results in terms of scan cost, which implies corresponding I/O bounds. We state these explicitly here for our final construction and following the theorems in the main text.

\paragraph{Part 1: Simple Catalan Hashing (Static, Low Load Factor).}
A naive way to store items inline in a linear-probing hash table would be to store each item \(x\) as a contiguous block of \(|x|\) words starting at or after its hash location \(h(x)\).
However, as observed by Thorup \cite{Thorup09}. this can blow up the cost of small items: a single large item of length \(L\) occupies \(L\) consecutive cells, so an item of length \(1\) that hashes into that interval is pushed \(\Omega(L)\) cells away from its hash location.

In Part~1 (\Cref{sec:catalan,sec:simple-catalan}), we introduce \emph{(Simple) Catalan hashing}, a static linear probing hash table (designed for low load factors $1 - \Theta(1)$) that resolves the problem described above. The basic idea is to store items non-contiguously, nesting smaller items inside larger ones. We then augment each cell of the table to store an $O(1)$-bit tag indicating where items begin and end. The items are laid out in such a way that, whenever a smaller item $x$ overlaps a larger one $y$, $x$ is contained \emph{entirely} within $y$. If one thinks of each tag as encoding either an open parenthesis (starting a new item) or a closing parenthesis (ending the current item), then the tags form a balanced-parentheses sequence (i.e., a \emph{Catalan sequence}~\cite{Stanley15Catalan}), hence the name of the data structure.

The main result of Part~1 is the following theorem:
\begin{restatable}[Simple Catalan Hashing]{theorem}{thmSimpleCatalan}
\label{thm:intro-simple}
Let $S$ be a static set of variable-length items stored in an $m$-cell table at load factor $\load(S) = \frac{1}{m}\sum_{y \in S} |y| \le 1 - \Omega(1)$. Simple Catalan hashing uses $O(1)$ bits of metadata per cell, answers positive queries for any item $x \in S$ with expected scan cost $O(|x|)$, and answers negative queries with expected scan cost $O(1)$.
\end{restatable}

Note that the bound for negative queries (queries to keys that are absent) does not even depend on the length $|x|$ of the item being queried. Negative queries make use of the following mechanism: If an item $x$ hashes to some cell $c$, and $x$ is the last item in the table with that hash, then at the end of $x$ we store an $O(1)$-bit tag that indicates this. A negative query for a key $k$ with hash $c$ can therefore stop as soon as it reaches the tag at the end of $x$. Of course, if $x$ is very long, then this could make the query very expensive! Critically, however, long items are necessarily rare (the sum of the lengths of all items is at most the size of the table), so a negative query is unlikely to share its hash with that of a long item. Summing over the items in the table, we show that the expected scan cost of a negative query is just $O(1)$.

\paragraph{Part 2: Write Optimization via Erasure Codes (Dynamic, Low Load Factor).}
Next, in Part~2, we turn to the question of how to support insertions (deletions are deferred to Part~5). Note that, here, we are still only focusing on hash tables that operate at low load factors---we consider higher load factors in Parts 4 and 5.  

Note that naively dynamizing the simple Catalan hash table results in severe \emph{write amplification} (insertions are very expensive). Suppose, for example, that the table stores a long item $y$ of length $L$, and we insert an item $x$ of length $1$ whose hash lands in the middle of $y$. In the static layout, $x$ interrupts $y$, so inserting $x$ requires shifting the remaining $\Omega(L)$ words of $y$ forward, giving the insertion a scan cost of $\Omega(L)$. Worse, if $y$ is itself nested inside an even longer item $y'$, then shifting $y$ forces $y'$ to shift as well, triggering a cascade of updates across items of increasing size.


In the external-memory literature, the classic way to resolve write amplification is to make the data structure \emph{write optimized}~\cite{ONeilChengGawlickONeil1996LSM,Arge03BufferTree,BrodalFagerberg2003LowerBounds,BenderEtAl2007StreamingBTrees,IaconoPatrascu2012ExternalDictionary,BenderFJJKPYZ15,ConwayFarachColtonShilane2018OptimalHashing}.
However, existing write-optimized data structures (such as $B^\eps$-trees~\cite{BrodalFagerberg2003LowerBounds}, LSM-trees~\cite{ONeilChengGawlickONeil1996LSM}, cache-oblivious streaming B-trees~\cite{BenderEtAl2007StreamingBTrees}, etc.) rely on hierarchical trees of buffers to batch small updates as they trickle down the tree. In a linear-probing hash table, however, there is no tree, and no natural way to add hierarchical buffers in a way that respects spatial locality. Thus, \emph{a priori}, write optimization would seem to be off the table.

Our solution is a new mechanism for write optimization in linear-probing hash tables: \emph{using erasure codes so that large items can themselves act as ``buffers'' for small items}. Instead of storing an item $y$ of length $s = |y|$ directly, we encode $y$ using an erasure  code $\Enc(y)$ of length $2s$ from which $y$ can be reconstructed using \emph{any} $s$ surviving symbols. When a newly inserted short item $x$ needs a cell that is currently occupied by a longer element $y$, $x$ does not move $y$ at all---it simply \emph{overwrites} $y$'s cell in place. As long as at least $s$ of the $2s$ cells of $\Enc(y)$ survive, any query for $y$ can still decode $y$ in a single contiguous scan, just as before. Only when half of $\Enc(y)$ has been overwritten do we \emph{reinflate} $y$ by reinserting it as a fresh length-$2s$ codeword; critically, at that point, at least $s = |y|$ words of shorter items have been written on top of $y$, which is enough to amortize the cost of reinflating $y$.

To illustrate the high-level idea behind this technique, we being in \Cref{sec:erasure-oracle} with a warmup in which we analyze this erasure-coded Catalan hash table under the simplifying assumption that an \emph{erasure oracle} notifies the algorithm whenever half of an item's codeword has been overwritten (and supplies the original item to be reinflated). Under this oracle assumption, we prove (\Cref{thm:oracle-model}) that positive queries and insertions both have amortized expected scan cost $O(|x|)$.

\paragraph{Part 3: Removing the Oracle (Dynamic, Low Load Factor).}
In Part~3 (\Cref{sec:backup-table}), we show how to remove the oracle assumption from Part~2. What makes removing the oracle nontrivial is that, when a short insertion overwrites a cell belonging to a long item $y$, the algorithm cannot, for example, afford to update any counter that is stored at the start of $y$. Because $y$ may be very long, the start of $y$ may lie far before the cell being overwritten, so updating a per-item counter on every overwritten cell would destroy our $O(|x|)$ bound on the scan cost of an insertion. This makes it difficult to identify \emph{when} an item $y$ has had more than half of its codewords overwritten.

Our solution is to do \emph{nothing at all} to $y$ when its cells are overwritten, and instead reinflate heavily overwritten items \emph{lazily during queries}. Whenever an item $x$ is inserted, we also store a backup copy of $x$ in a secondary \emph{backup table} $\mathcal{R}$ that is interleaved into the same physical array. Most queries for an item $y$ simply find at least $|y|$ surviving symbols of $\Enc(y)$ in the primary table and decode $y$ with scan cost $O(|y|)$. If, however, more than half of $\Enc(y)$ has been overwritten, the query falls back to recovering $y$ from the backup table $\mathcal{R}$ and reinflates $\Enc(y)$ in the primary table.

At first glance, introducing $\mathcal{R}$ might appear to be circular, since $\mathcal{R}$ must itself dynamically store variable-length items. The crucial difference is that a query for $y$ only accesses $\mathcal{R}$ when at least $|y|$ cells of $\Enc(y)$ have been overwritten since $y$ was last inserted or reinflated. We can therefore \emph{amortize} the cost of querying $\mathcal{R}$ (and reinflating $y$) against the insertions that caused those overwrites. Critically, this means that we can get away with $\mathcal{R}$ supporting slower queries than what we want in the full data structure---it just needs to support all operations with scan cost $O(|x|\log^{1+\delta}|x|)$. The final contribution of \Cref{sec:backup-table} is to construct a backup table that does just that. Integrating this into the full data structure gives the following theorem:

\begin{restatable}[Dynamic Catalan Hashing at Low Load Factor]{theorem}{thmDynamicLow}
\label{thm:intro-dynamic-low}
Fix any constant $\delta > 0$. There is an insertion-only dynamic Catalan hash table that supports some constant load factor $\Omega(1)$, and that answers positive queries for any item $x$ with amortized expected scan cost $O(|x|)$, answers negative queries with expected scan cost $O(1)$, and supports insertions of any item $x$ with amortized expected scan cost $O(|x| \log^{1+\delta} |x|)$.
\end{restatable}

\paragraph{Part 4: Fair Catalan Hashing (Static, High Load Factors).}
In Part~4 (\Cref{sec:fair-catalan}), we move from constant load factors to the \emph{high-load-factor regime} (load factor $1 - \eps$ for some small $\eps \in (0, 1)$). To tackle this regime, we begin by returning to the static setting.

At high load factors, simple Catalan hashing ends up encountering the same quadratic barrier as classical linear probing---the worst-case expected scan cost for an item $x$ becomes $\Theta(\eps^{-2} |x|)$. The standard fix in the unit-size setting is to use the Robin Hood hashing~\cite{AmbleKnuth74,CelisLarsonMunro1985RobinHood}, which sorts elements in each run in increasing order of hash. But, for variable-length items, this approach fails, because a single long item can blow up the query time of many subsequent short items with larger hashes.

To overcome this obstacle, we establish a connection between variable-length linear probing and \emph{preemptive single-server scheduling}~\cite{Schrage1968SRPT,Kleinrock67,Kleinrock76,BenderCM98,BansalHarcholBalter2001SRPTUnfairness,FriedmanH03,HarcholBalter13}. If we view each item $x$ as a job of size $|x|$ arriving at time $h(x)$, then we can reinterpret the task of placing the items into the table as a scheduling problem. Under this view, the scan cost of an item $x$ divided by its size $|x|$ is the same as the \emph{stretch} of the corresponding job.

What we want, then, is a scheduling policy that is \emph{fair} across all job sizes, guaranteeing worst-case expected stretch $O(\eps^{-1})$. In queuing theory, the simplest such policy is \emph{last-in-first-out with preemption}, in which a newly-arrived jobs immediately preempt any existing jobs. In our hash-table setting, this policy can be viewed as giving priority to items with \emph{larger} hashes. Note that this prioritization is the exact \emph{opposite} of the classical Robin Hood ordering \cite{AmbleKnuth74,CelisLarsonMunro1985RobinHood}. We call the resulting data structure \emph{anti-Robin-Hood Catalan hashing (ARCH)}. 

In \Cref{sec:fair-catalan}, we give a simple, self-contained proof that the ARCH table does, indeed, guarantee an expected scan length of at most $\eps^{-1} |x|$.\footnote{It is worth noting that this bound holds not just asymptotically but exactly.} Perhaps ironically, this means that the \emph{anti-}Robin-Hood ordering is actually fairer than its Robin Hood counterpart, once we allow for variable-length items. 

\begin{restatable}[Anti-Robin-Hood Catalan Hashing]{theorem}{thmARCH}
\label{thm:arch}
Let $\eps \in (0, 1)$, and let $S$ be a static set of items stored in an $m$-cell table with load factor $\load(S) \le 1 - \eps$. Anti-Robin-Hood Catalan hashing (ARCH) uses $O(1)$ bits of metadata per cell, answers positive queries for any item $x \in S$ with expected scan cost at most $|x|/\eps$, and answers negative queries with expected scan cost at most $1 + 1/\eps$.
\end{restatable}

We remark that, although the anti-Robin-Hood ordering bounds the expected stretch of every item, which is what we need in Part~5, its expected \emph{maximum} stretch can be quite large: if the table contains many items of length $2$ and a single item of length $L$, then with constant probability the long item preempts one of the short items, giving that item stretch $\Omega(L)$. As an additional result, we show how to bound the expected maximum stretch as well. We adapt the \emph{fair-sojourn scheduling protocol} of Friedman and Henderson~\cite{FriedmanH03} to design a variant of Catalan hashing with expected scan cost at most $1+|x|/\eps$ and expected maximum stretch at most $1+(\log n+1)/\eps$ (\Cref{thm:fsch}). 

\paragraph{Part 5: Dynamic Catalan Hashing at High Load Factors.}
Finally, in Part~5 (\Cref{sec:dynamic-high-load}), we combine our static ARCH construction from Part~4 with our dynamic low-load-factor construction from Parts~2 and~3 to obtain a fully dynamic Catalan hash table at load factor $1 - \eps$ (supporting both insertions and deletions).

Note that the dynamic erasure-coded hash table from Parts~2 and~3 cannot be run directly at load factor $1 - \eps$: its write-optimization machinery inherently loses a constant factor in space.
Conversely, the scheduling-based approach from Part~4 achieves load factor $1 - \eps$ with only $O(1)$ bits of metadata per cell, but is inherently static.

To get the best of both worlds, we use a structure inspired by \emph{graveyard hashing}~\cite{BenderKuszmaulKuszmaul2021GraveyardHashing}. At a high level, we store almost all of the items in a static ARCH table $\HT_{\mathrm{full}}$ at load factor $1 - O(\eps)$, while routing new insertions into a much smaller dynamic erasure-coded table $\HT_{\mathrm{small}}$ of size $\Theta(\eps m)$ that operates at constant load factor. Because $\HT_{\mathrm{small}}$ is only a $\Theta(\eps)$ fraction of the size of $\HT_{\mathrm{full}}$, its constant-factor space overhead consumes only $O(\eps m)$ cells in total. Moreover, so that an operation that accesses both tables still scans just one small region of memory, we \emph{interleave} $\HT_{\mathrm{small}}$ uniformly into $\HT_{\mathrm{full}}$ within a single physical array---placing one slot of $\HT_{\mathrm{small}}$ every $\Theta(\eps^{-1})$ slots of $\HT_{\mathrm{full}}$---and use a shared hash function $h(x)$ for both tables. That way, any operation that probes both $\HT_{\mathrm{full}}$ and $\HT_{\mathrm{small}}$ still traverses a single contiguous region of memory around $h(x)$ (stretching the physical scan cost of $\HT_{\mathrm{small}}$ by a factor of $\Theta(\eps^{-1})$).

Insertions go directly into $\HT_{\mathrm{small}}$, and a query simply searches both tables. After every $\Theta(\eps m)$ words of insertions, we perform a global rebuild that merges $\HT_{\mathrm{small}}$ into $\HT_{\mathrm{full}}$ using $O(1)$ sequential passes over the table. Because the rebuild occurs only once every $\Theta(\eps m)$ words of updates, its $O(m)$ cost amortizes to $O(\eps^{-1})$ per inserted word---or $O(|x|/\eps)$ for an item $x$. To support deletions without making negative queries depend on the key length, we keep constant-size fingerprints of deleted static keys in another interleaved table, checking for collisions during rebuilding. The details, together with the low-load deletion mechanism, appear in \Cref{sec:dynamic-high-load} and \Cref{app:deletions}. Putting these pieces together gives our main result:

\begin{restatable}[Dynamic High-Load Catalan Hashing]{theorem}{thmDynamicHigh}
\label{thm:intro-dynamic-high}
Let $\eps \in (0, 1/2)$ and let $\delta > 0$ be a constant. There is a dynamic Catalan hash table operating at load factor $1 - \eps$ that supports positive queries for any item $x$ with amortized expected scan cost
\[
    O\left(\frac{|x|}{\eps}\right),
\]
supports negative queries with amortized expected scan cost $O(1/\eps)$, and supports insertions and deletions of any item $x$ with amortized expected scan cost
\[
    O\left(\frac{|x| \log^{1+\delta}|x|}{\eps}\right).
\]
\end{restatable}

As an immediate corollary we also get the following result in the external memory model, establishing the existence of a linear-probing-like data structure for items of arbitrary sizes. 

\begin{restatable}[Dynamic High-Load Catalan Hashing in External Memory]{corollary}{corDynamicHighEM}
\label{cor:intro-dynamic-high-em}
Suppose $M \ge C_M(B + s_{\max}\eps^{-2}\log^3 m)$ for a sufficiently large constant $C_M$, where $s_{\max}$ is the maximum length of any item stored or inserted during the operation sequence. The hash table of \Cref{thm:intro-dynamic-high} is cache-oblivious and, for every block size $B$, supports positive queries for any item $x$ in amortized expected
\[
    1 + O\left(\frac{|x|}{\eps B}\right)
\]
I/Os, supports negative queries in amortized expected $1 + O(1/(\eps B))$ I/Os, and supports insertions and deletions of any item $x$ in amortized expected
\[
    1 + O\left(\frac{|x| \log^{1+\delta}|x|}{\eps B}\right)
\]
I/Os.
\end{restatable}

\paragraph{Paper Outline.} The rest of the introduction is spent discussing in more detail related work on linear probing and on external-memory data structures. Then, after preliminaries in \Cref{sec:preliminaries}, we introduce the basic machinery that all of our static constructions share in \Cref{sec:catalan}, and then we complete Parts 1--5 in \Cref{sec:simple-catalan,sec:erasure-oracle,sec:backup-table,sec:fair-catalan,sec:dynamic-high-load}, respectively.

\subsection{Additional Related Work}
For fixed-size items, linear probing has been studied for more than half a century. The first analysis is due to Knuth in 1963~\cite{Knuth63note}, who showed that, when the hash table is $1 - \eps$ full, insertions and negative queries take $\Theta(\eps^{-2})$ expected time; Knuth's analysis was never published, and the first published analysis was in 1966 by Konheim and Weiss~\cite{KonheimWeiss1966Occupancy}. Amble and Knuth~\cite{AmbleKnuth74} showed that, by keeping the items in each run in sorted order, one can reduce the worst-case expected query time to $O(\eps^{-1})$ (this has become known as Robin-Hood hashing~\cite{CelisLarsonMunro1985RobinHood} due to the way in which it gives priority to elements which have traveled farther). Much more recently, Bender, Kuszmaul, and Kuszmaul~\cite{BenderKuszmaulKuszmaul2021GraveyardHashing} showed how to also achieve this $O(\eps^{-1})$ bound for insertions and deletions; this is achieved by \emph{graveyard hashing}, in which one occasionally rebuilds the hash table to have strategically placed \emph{gaps} (tombstones) that break apart runs for subsequent insertions. The authors also showed \cite{BenderKuszmaulKuszmaul2021GraveyardHashing,BravermanKuszmaul2025LinearProbing} that, even in standard ordered linear probing with tombstone-based deletions, the amortized expected insertion time is better than one might assume, clocking in at $\Theta(\eps^{-1} \log^{1.5} \eps^{-1})$. Most recently, there has been interest in what the optimal scan cost is in a (insert-only) hash table that never moves items after they are inserted. Zamir~\cite{Zamir2026Locality} showed that, in this setting, the $\Theta(\eps^{-2})$ scan cost of linear probing is, in a certain sense, optimal: no such hash table can achieve scan cost $o(\eps^{-2})$ at all load factors simultaneously. In concurrent work, Krapivin, Kuszmaul, and Wang~\cite{KrapivinKuszmaulWang2026NonGreedy} showed that, if $\eps$ is known in advance, then it is possible to achieve a worst-case expected scan cost for insertions and positive queries of $O(\eps^{-1} \log \eps^{-1})$, and that (somewhat surprisingly) no strategy can do better than $\Omega(\eps^{-1} \sqrt{\log \eps^{-1}})$. All of this work, however, focuses on the setting in which items have a fixed size.

All of the data structures in this paper come with natural corollaries for the external-memory model~\cite{AggarwalVitter1988IO} (see \Cref{sec:preliminaries} for a more formal treatment of this model). In many settings, the algorithm may not know the block-size parameter $B$ in advance (or may even be interacting with multiple different types of caches, using different block sizes, simultaneously). Algorithms that get strong guarantees even in this setting (such as those in this paper) are known as \emph{cache oblivious}~\cite{FrigoLPR12}. These models underlie a large body of search structures and write-optimized dictionaries, including cache-oblivious B-trees~\cite{BenderDemaineFarachColton2005COBTrees}, packed-memory arrays and their adaptive variants~\cite{BenderHu2007AdaptivePMA}, $B^\eps$-trees~\cite{BrodalFagerberg2003LowerBounds}, log-structured merge (LSM) trees~\cite{ONeilChengGawlickONeil1996LSM}, and cache-oblivious streaming B-trees~\cite{BenderEtAl2007StreamingBTrees}. There has also been a great deal of work on designing hash tables and unordered dictionaries in this setting \cite{FaginNievergeltPippengerStrong1979ExtendibleHashing,Litwin1980LinearHashing,Larson1988DynamicHashTables,JensenPagh2008ExternalHashing,PaghWeiYiZhang2014CacheObliviousHashing,IaconoPatrascu2012ExternalDictionary,ConwayFarachColtonShilane2018OptimalHashing,BenderKuszmaulKuszmaul2021GraveyardHashing,BenderCFKT23Iceberg,BravermanKuszmaul2025LinearProbing}, but with a focus on settings where all items have the same lengths.

Prior to this paper, most of the work on external-memory data structures for variable-length objects has focused on how to extend the $B$-tree and its variants to the variable-length setting \cite{FerraginaGrossi1999StringBTree,BrodalFagerberg2006StringDictionaries,BenderFarachColtonKuszmaul2006StringBTrees}. In contrast, in the current paper, we study the question of how to best extend the guarantees of a \emph{hash table}. Perhaps surprisingly (with the exceptions of \cite{Thorup09,KurpiczLS23} which we will discuss shortly), this question does not appear to have been widely studied in prior work.

The idea of placing variable-length strings into a linear-probing hash table appears to have been first discussed in Thorup's paper on designing hash functions for hash tables storing strings~\cite[Section~6, Theorem~6.1]{Thorup09}. For load factor $\alpha \le 0.9$, Thorup bounds the expected probe length for a positive query to a string $x$ by $|x| + O(\alpha R)$, where $R = (\sum_{y \in S} |y|^2) / (\sum_{y \in S} |y|)$ is the second-moment-to-first-moment ratio of the string lengths in the table $S$. When the table contains even a few very long strings, $R$ can be much larger than $|x|$, causing queries for short strings to pay for the lengths of unrelated long strings. One of the first insights in the current paper, is that by allowing shorter strings to nest within longer ones (via Catalan hashing), we can avoid this kind of cost blow-up for queries to short strings.

Finally, as noted earlier, another natural approach to storing variable-length strings in external memory is to use a layer of indirection, storing an internal-memory index that keeps track of where each object resides in external memory. Naively, such a solution must incur $\Omega(\log n)$ bits of internal-memory metadata per element. PaCHash \cite{KurpiczLS23}, however, shows that it is possible to build a more sophisticated version of this approach that stores only $2 + \log a + o(1)$ bits of internal-memory metadata per \emph{block} of external-memory storage, and that offers queries in expected $1 + O(|x|/B) + 1/a$ I/Os (where $a$ is a parameter of our choice). For settings where the data is static, where $B$ is known in advance, and where there is sufficient internal memory, this solution has been shown to perform well in practice \cite{KurpiczLS23}.

Throughout this paper, we assume access to fully random hash functions. There has also been a great deal of work on how to implement linear probing with explicit hash functions \cite{PaghPR09,PatrascuThorup2016KIndependence,Thorup09,ThorupZhang2012TabulationFiveIndependent,PatrascuThorup2012SimpleTabulation,BerceaEtAl2023LocallyUniform}, as well as, more generally, how to simulate random hash functions for essentially any hash table \cite{Siegel2004UniversalClasses,PaghPagh2008UniformHashing,DietzfelbingerRink2009SplittingTrick} (see, e.g., the discussion in \cite{BenderFKKL22}).



\section{Preliminaries}
\label{sec:preliminaries}

\paragraph{Conventions.}
All logarithms are base $2$. In asymptotic cost bounds, we use the convention $\log s=\max\{1,\log_2 s\}$ for $s\ge1$.
For ease of exposition, we ignore divisibility and rounding issues whenever they do not affect the stated asymptotic bounds.

\paragraph{Items, Keys, and Lengths.}
All lengths in this paper are measured in \defn{words}. As in the standard word-RAM model, we assume that a word has at least $c_w \log m$ bits for some constant $c_w$.

The elements stored in our hash table are variable-length \defn{items}. Each item $x = (k, v)$ consists of a \defn{key} $k$ and a \defn{value} $v$, both of which may have variable length (the value may also be empty). We store $x$ as a sequence of $|x| \ge 1$ words, which we call the \defn{length} (or \defn{size}) of $x$, encoded so that the key can be read off from a prefix of $x$ (for example, by storing the length of $k$ at the start of $x$, or by ending $k$ with a terminator, as in a C string). 
Each key belongs to at most one item in the hash table, and we sometimes write $k(x)$ for the key of $x$ and \(v(x)\) for the value of \(x\). 

A \defn{query} is given only a key $k$; it returns \textsc{Found} and the value $v(x)$, where \(x\in S\) is the item with key \(k\) if there is one (a \defn{positive} query), and \textsc{Not Found} otherwise (a \defn{negative} query). (If values are empty, a positive query simply returns \textsc{Found}).
In particular, a query does not necessarily know the length of the item it is looking for.
In our query bounds, for positive queries, $x$ denotes the item found.
For negative queries, the bounds hold with \(|x| = 1\), independently of the length of the query key.
All hash functions are applied to keys, so the hash $h(x) = h(k(x))$ of an item depends only on its key, and we write $h(k)$ where convenient.

\paragraph{Hashing, Load Factor, and Runs.}
In this paper, a hash table $\HT$ consists of $m$ cells, indexed cyclically as $\{0, 1, \ldots, m-1\}$ (or equivalently $\Z / m\Z$). We assume access to a fully random hash function $h$ that maps each key independently and uniformly to the continuous circle $[0, m)$.
We treat hash values as real numbers, and assume that basic operations on them, such as arithmetic and comparisons, take $O(1)$ time.\footnote{We make this assumption for ease of exposition. To break ties with high probability, we only need \(O(\log m )\) bits of precision.}
We also assume that the sequence of operations is fixed in advance, independently of the hash functions, or equivalently, that the input sequence is chosen by an \defn{oblivious adversary}.
The \defn{canonical cell} of an item $x$ is determined by its key \(k(x)\), defined as $c(x) = c(k(x)) = \lfloor h(x) \rfloor \in \{0, \ldots, m-1\}$.
The fractional part of $h(x)$ may be used to break ties between items hashing to the same canonical cell.

Given the set of items $S$ stored in an $m$-cell table, the \defn{total volume} of $S$ is $\Vol(S) = \sum_{x \in S} |x|$, and the \defn{load factor} (or \defn{fill}) of the table is
\(
    \load(S) = \Vol(S)/m
\).
When $\load(S) \le \alpha < 1$, we say that the table has load factor at most $\alpha$ and \defn{slack} $\eps = 1 - \alpha$. 
Any metadata used by the algorithm is not counted in the load factor, but the metadata used in the hash table in this paper only adds a constant factor overhead in the constant-load regimes and is a low-order term in the high-load regimes.
In the low-load-factor sections of the paper (\Cref{sec:simple-catalan,sec:erasure-oracle,sec:backup-table}), we assume $\alpha \le 1 - \Omega(1)$, i.e., the load factor is bounded away from $1$ by a positive constant. 
In the high-load-factor sections (\Cref{sec:fair-catalan,sec:dynamic-high-load}), we parameterize the performance directly in terms of $\eps$.

A \defn{run} (or \defn{cluster}) is a maximal contiguous interval of occupied cells in $\HT$. Whenever $\load(S) < 1$, at least one cell of $\HT$ is empty, so every run is preceded immediately by an empty cell and has a well-defined left-to-right linear order obtained by unwrapping the circular array at that empty cell. For a query to the key of an item $x$, the \defn{scan length} $L_x$ is the number of contiguous cells inspected by the query starting from $c(x)$, and the \defn{query stretch} is $L_x / |x|$.

\paragraph{Scan Cost.}
Nearly all of our algorithms access a contiguous cyclic interval of the hash table.
Following classical linear-probing work~\cite{Knuth63note,Knuth98}, we define the \defn{scan cost} of an operation on the hash table as the size of the smallest cyclic interval that contains all of its accessed cells.
Our primary analysis will report the scan cost of our algorithms, which then can be easily converted into external-memory cost (defined below).

\paragraph{The External-Memory and Cache-Oblivious Models.}
We use the standard cache-oblivious model of Frigo, Leiserson, Prokop, and Ramachandran~\cite{FrigoLPR12}, building on the external-memory (I/O) model of Aggarwal and Vitter~\cite{AggarwalVitter1988IO}. 
In the external-memory model, memory consists of an internal cache of size $M$ words and an external memory partitioned into blocks of $B$ consecutive words.
An \defn{I/O} transfers one block of $B$ words between cache and external memory. The cost of a sequence of operations is the number of I/Os that it incurs. 
An algorithm is \defn{cache-oblivious} if its specification does not depend on the parameters \(M\) and \(B\); these appear only in the analysis.

We store the table using a uniformly random rotation (chosen at initialization, independently of the hash function).
A forward scan of a contiguous interval of $L$ cells costs $O(1+L/B)$ I/Os, and $1+O(L/B)$ in expectation over the rotation.
The same bounds hold for an operation making $O(1)$ forward passes over the interval.
We refer to this cost as the \defn{scan property}.

For ease of exposition, we assume circularity of the array in the remainder of this paper and do not explicitly refer to circular order, cyclic intervals, etc., except where necessary.
We will sometimes implicitly or explicitly refer to the hash or other functions as \defn{interpreted cyclically} as follows.
If a run starts at cell \(a\), wraps around the end of the array and ends at cell \(b\) and \(h(x)\in [0,b)\), we consider \(x\)'s hash to be \(h(x)+m\), so that within the run, hashes are always monotonic.

We remark that one can also straightforwardly optimize the data structures in this paper to be efficient in the word-RAM model, so that the time per operation is the same as the scan cost (plus $O(|x|)$). The only nontrivial optimization needed is to replace the erasure codes introduced in \cref{sec:erasure-oracle} with codes that supports linear-time operations \cite{alon1996linear}.

\paragraph{Cache-Size Assumptions.}
We make what we call the \defn{working-set assumption}, which says that the cache has room for the query key, one other key, the parser state, and a sufficiently large constant number of blocks. This is the only cache-size assumption needed by the static constructions (\Cref{sec:simple-catalan,sec:fair-catalan}).
The dynamic constructions will require moderate additional amounts of memory as stated in their respective sections.

\section{Catalan Hashing}\label{sec:catalan}
This section introduces the common infrastructure used by the hash tables in this paper, all of which we refer to as types of \defn{Catalan hashing}. The static constructions in this paper (\Cref{sec:simple-catalan,sec:fair-catalan}) will make direct use of the framework described here, while the dynamic sections (\Cref{sec:erasure-oracle,sec:backup-table,sec:dynamic-high-load}) will need to white-box extend it in various ways.

\paragraph{The Nesting Property.}
In a Catalan hash table, we do not necessarily store each item contiguously. Instead, we allow some items to \emph{preempt} others. For example, if we have one very large item $y$, say of length $100$ with canonical cell $c(y) = 1$, and one very small item $x$, say of length $2$ with $c(x) = 3$, then we might place the first two words of $y$ in cells $1$ and $2$, then place the two words of $x$ in cells $3$ and $4$, and then resume placing the remaining $98$ words of $y$ in cells $5$ through $102$. This allows the short item $x$ to avoid paying any kind of substantial query penalty for being placed near a large item $y$. 

Formally, our static constructions will satisfy the following nesting property. For each stored item $x$ in a run, let $I_x = [a_x, b_x]$ denote the interval of cells from the first word of $x$ ($a_x$) to the last word of $x$ ($b_x$).
\begin{definition}[Catalan Nesting Property]
\label{def:catalan-property}
A placement of items into $\HT$ satisfies the \defn{Catalan nesting property} if, within every run, for any two distinct items $x$ and $y$:
\begin{enumerate}[label=(\arabic*)]
    \item Either $I_x$ and $I_y$ are disjoint ($I_x \cap I_y = \emptyset$), or one is strictly contained in the other ($I_y \subsetneq I_x$ or $I_x \subsetneq I_y$); and
    \item Whenever $I_y \subsetneq I_x$, no cell in $I_y$ stores a word of $x$ (equivalently, the entire interval $I_y$ lies strictly between two consecutive cells assigned to $x$).
\end{enumerate}
\end{definition}

When describing a given static construction (in \Cref{sec:simple-catalan,sec:fair-catalan}), the key distinguishing feature will be the \emph{placement rule} that we use within the table. This rule is responsible for deciding which items are nested within which other items, and where the nesting starts and ends. The simplest construction that we consider (simple Catalan hashing in \Cref{sec:simple-catalan}) uses a placement rule in which smaller items always take priority over larger items. This, as we will see, leads to good results in hash tables that are not very full (load factor $1 - \Omega(1)$), but not for hash tables that are much fuller (load factor $1 - \epsilon$). To obtain hash tables that behave well even under high load, we will end up needing a more sophisticated placement rule (fair Catalan hashing in \Cref{sec:fair-catalan}).

\paragraph{The Catalan Representation.}\phantomsection\label{sec:catalan-rep}
The static constructions in \Cref{sec:simple-catalan,sec:fair-catalan} use the following basic metadata scheme which we call the \defn{Catalan representation}. 

The Catalan representation augments each cell of $\HT$ with a $3$-bit delimiter tag taking one of five values:
\[
    \copen{}, \quad \cmiddle{}, \quad \cclose{}, \quad \copenclose{}, \quad \cempty{}.
\]
An unoccupied cell is marked \cempty{}. If an item $x$ has length $|x| = 1$, the single cell holding its word is marked \copenclose{}. If $|x| \ge 2$, the cell holding the first word of $x$ is marked \copen{}, the cell holding the last word of $x$ is marked \cclose{}, and the $|x|-2$ cells holding the interior words of $x$ are marked \cmiddle{}.

The name \emph{Catalan hashing} comes from the combinatorial structure of these delimiters: in any run satisfying \Cref{def:catalan-property}, if we ignore \cmiddle{} tags and view \copen{} as ``\texttt{(}'', \cclose{} as ``\texttt{)}'', and \copenclose{} as ``\texttt{()}'', the sequence of delimiters in the run forms a balanced parenthesis string, known as a Catalan sequence~\cite{Stanley15Catalan}.

Each cell additionally stores $2$ bits called the \defn{home} and \defn{stop} bits. The home and stop bits indicate home and stop cells, which are used to limit the scan length of negative queries.
A cell which is the canonical cell for some item in the table is called a \defn{home cell}.
Each home cell \(j\) has a corresponding \defn{stop cell}, which is the last cell containing a word from an item whose canonical cell is \(j\).

\paragraph{Parsing the Nested Structure.}\phantomsection\label{sec:catalan-parser}
For a placement satisfying the Catalan nesting property, any occupied cell belongs to the \emph{innermost} item interval $I_x$ that contains it.
This makes it possible for a query starting at any cell $t$ (including a cell in the middle of a run) to parse the stream of cells using a stack-based pushdown automaton which we will refer to as a \defn{Catalan parser}.
At a high level, the parser maintains a stack of the items that it has seen open (via their \copen{} tags) but not yet close (via their \cclose{} tags). By the nesting property, every \cmiddle{} or \cclose{} cell that it reads belongs to the item on top of the stack, or, if the stack is empty, to an item that opened before $t$. For each item on the stack, the parser tracks how many of the item's words it has read and whether they agree with the query key so far, so that it can recognize a matching item as soon as that item's last word is read. The parser's behavior is described in more detail in \Cref{app:parser}.

\paragraph{Query Procedure.}\phantomsection\label{sec:catalan-query}
With the help of this parser, queries work as follows. A query for a key $k$ runs the Catalan parser forward from the canonical cell $c(k)$, and stops as soon as one of the following occurs:
\begin{itemize}
    \item $c(k)$ is not a home cell (its home bit is unset). Then no item in the table has canonical cell $c(k)$, so the query returns \textsc{Not Found} without scanning further.
    \item The parser finishes reading an item $x$ with $k(x) = k$. The query returns \textsc{Found}, together with the value of $x$.
    \item The parser finishes reading an item $y$ with $k(y) \neq k$ and $c(y) = c(k)$ whose last cell is a stop cell (its stop bit is set). Every item with canonical cell $c(k)$ ends at or before this cell, so the query returns \textsc{Not Found}.
\end{itemize}

\paragraph{Negative versus Positive Queries.}
One risk for negative queries is that, if an item $y$ is not in the table, but has the same canonical cell $c(y) = c(x)$ as a very large item $x$ in the table, then the scan cost for $y$ will be very high (going at least to the end of $x$). On the other hand, large items $x$ are rare, since the sum of the sizes of all items can be at most $m$. This is formalized by the following lemma, which shows that efficient positive queries immediately imply efficient negative queries, regardless of the placement rule used.

\begin{lemma}[Negative Queries Versus Positive Queries]
\label{lem:stop-marks}
Consider a static implementation of Catalan hashing using an arbitrary placement rule. For each item $x \in S$, let $L_x$ be the scan cost for querying $x$. Let $C$ be a parameter, and suppose that $\E[L_x] \le C|x|$ for each $x \in S$. Then, for any item $y$ \emph{not} in the table, the expected scan cost of a negative query for $y$ is at most $1 + C$. 
\end{lemma}
\begin{proof}
Let $D$ be the scan cost of the negative query for $y$. We first observe that
\begin{equation}\label{eqn:negative-vs-positive}
    D \le 1 + \sum_{x \in S} L_x \cdot \ind[c(x) = c(y)].
\end{equation}
Indeed, if $c(y)$ is not a home cell for any $x \in S$, then the query stops immediately and $D = 1$; otherwise, $D$ is simply $\max_{x \in S : c(x) = c(y)} L_x$. Either way, \eqref{eqn:negative-vs-positive} holds.

We may therefore conclude that 
\[
    \E[D] \le 1 + \sum_{x \in S} \E[L_x \cdot \ind[c(x) = c(y)]].
\]
Since $c(y)$ is uniformly random, and independent of the randomness used to determine $L_x$ (or more generally, to place the items $S$ into the table), we have that $\E[L_x \cdot \ind[c(x) = c(y)]] = \E[L_x]/m$. Therefore,
\[
    \E[D] \le 1 + \frac{1}{m} \sum_{x \in S} \E[L_x] \le 1 + \frac{C}{m} \sum_{x \in S} |x|.
\]
Since $\sum_{x \in S} |x| \le m$, the claimed bound of $\E[D] \le 1 + C$ follows.
\end{proof}

\paragraph{Priority Rules and Construction.}
All of the static placement rules in this paper are derived from different priority rules.
A \defn{priority rule} assigns a priority \(\pi_x\) to each item \(x\in S\), which can potentially depend on all the items in \(S\) and their hash values.
When multiple items contend for a given cell, a word from the item of highest priority is placed there.
We write \(x \succ y\) to denote \(\pi_x > \pi_y\).
A priority rule uniquely determines the placement rule, which we realize with the following equivalent constructions.

\begin{description}[labelindent=1em, leftmargin=2.5em, style=sameline]
\item[Bottom-up:] Place the items in decreasing priority order as follows.
For each item \(x\), place the words of \(x\) in the first \(|x|\) empty cells at or after \(c(x)\).
Items are never moved, so the previous placement of higher priority items restricts the empty cells available to $x$.

\item[Left-to-right:]
Starting from beginning of the array, scan forward, maintaining a max-heap of items by priority order and processing cells one at a time.
For each cell \(c\), insert all items \(x \in S\) with \(c(x) = c\) into the heap, then place the next word of the maximum-priority item into \(c\).
When all the words of an item in the heap are placed, extract it.
At the end of the table, if there are still words in the heap left to place, restart at the beginning.
This second pass evicts the word in each cell if it has lower priority than the maximum-priority item in the heap, in which case the evicted word is inserted back into the heap.
Because the load of \(S\) is less than 1, the heap will empty before the end of the second pass.

\item[Arbitrary order:] Place the items in \(S\) in an arbitrary order.
Call a cell \defn{available} for \(x\) if it is empty or contains a lower-priority item.
For each item \(x\), place the words of \(x\) in the first \(|x|\) available cells after \(c(x)\).
Whenever a word of a lower-priority item $y$ is evicted, reinsert that word and the remaining suffix of $y$ in logical word order.
Starting at the evicted cell, use the first cells that are empty or contain words of $y$ or of a lower-priority item, so that the suffix shifts to the right while reusing cells already assigned to $y$.
Apply the same rule to any further evictions, continuing until all words are placed.
This process must terminate, since the load of the table is less than $1$.
\end{description}

A placement rule determined by a priority automatically satisfies the Catalan nesting property.

\begin{lemma}\label{lem:priority-nesting}
    If the placement of items in \(\HT\) is determined by a priority rule, then it satisfies the Catalan nesting property.
\end{lemma}

\begin{proof}
    Let \(x \succ y\in S\).
    In the bottom-up construction, all the words of \(x\) are placed first in the first \(|x|\) free cells.
    Therefore, there are no free cells in \(I_x\), and so when \(y\) is placed, no words of \(y\) will be placed in \(I_x\).
\end{proof}

Note that for a \emph{static} construction for a Catalan hash table, all we need to specify is the priority rule used. This is captured formally via the following definition, which we invoke in \Cref{sec:simple-catalan,sec:fair-catalan}.
\begin{definition}\label{def:catalan}
    Given a priority rule \(\pi\), the \defn{static Catalan hash table} \(\HT_\pi\) is the hash table with placement rule and construction determined by \(\pi\), using the Catalan representation, parser and query procedure.
\end{definition}

\section{Simple Catalan Hashing (Static, Low Load)}
\label{sec:simple-catalan}

We now present and analyze simple Catalan hashing, a static implementation of Catalan hashing designed to operate at low load factors.

\paragraph{Prioritizing Smaller Items.}
Simple Catalan hashing prioritizes items according to their length: shorter items have strictly higher priority than longer items.
Ties between items of the same length are broken in favor of the item with the smaller hash value.
Formally, $\pi(y) > \pi(z)$ if $|y| < |z|$, or if $|y| = |z|$ and $h(y) < h(z)$.

Given \(S\) with \(\load(S) < 1\), we define the simple Catalan hash table to be \(\HT_\pi\) as in \Cref{def:catalan}.
Following \Cref{sec:catalan}, this defines a placement rule and construction, and uses the Catalan representation and query algorithm.

\paragraph{Analysis. }
A key property of simple Catalan hashing is that, to analyze the cost of a positive query for an item $x$ of length $\ell = |x|$, we can ignore all items in $S$ of length greater than $\ell$.

Formally, let $Y = \{y \in S \mid |y| \le \ell\}$ be the subset of items of length at most $\ell$, and imagine forming a Simple Catalan hash table $\HT_Y$ containing \emph{only} the items in $Y$. In $\HT_Y$, every cell that is empty corresponds in the full table to either to an empty cell or to a cell occupied by some item $z \in S \setminus Y$ of length $|z| \ge \ell + 1$. 

\begin{lemma}\label{lem:simple-domination}
    The cost to query $x$ in \(\HT_Y\) is the same as the cost to query $x$ in \(\HT\).
\end{lemma}
\begin{proof}
    In the bottom-up construction, the items of $Y$ are placed before every item of $S \setminus Y$, so each item of $Y$ occupies exactly the same cells in $\HT$ as in $\HT_Y$. Since $x \in Y$, the lemma follows. 
\end{proof}

We now bound the length of the run containing \(c(x)\) in \(\HT_Y\). We also prove a bound on the second moment of the run length, which we will use later in the paper.

\begin{lemma}\label{lem:simple-run-length}
    Let \(W_x\) be the length of the run in \(\HT_Y\) containing \(c(x)\).
    Then
    \[
        \E[W_x] = O(|x|) \quad\textrm{and}\quad \E[W_x^2] = O(|x|^2).
    \]
\end{lemma}

\begin{proof}
Condition on \(c(x)\), and partition the table into blocks of \(|x|\) cells.
Let \(I\) be the set of blocks that intersect the run containing \(c(x)\), and let \(Q = |I|\).
Every word in the run belongs to an item hashing into the run. All of the blocks in \(I\) except for the two end blocks are entirely covered by the run, and \(x\) itself accounts for only \(|x|\) words, so the items of \(Y\setminus\{x\}\) hashing into \(I\) have total length at least
\[
    \max\{(Q-3)|x|,\, 0\}.
\]

For a fixed interval \(J\) of \(q\) blocks, let \(T(J)\) be the total length of the items of \(Y \setminus \{x\}\) hashing into \(J\).
Then \(T(J)\) is a sum of independent random variables, one per item of \(Y \setminus \{x\}\), each taking values in \([0, |x|]\), and
\[
    \E[T(J)] \le \sum_{y \in S} |y| \cdot \frac{q|x|}{m} \le \alpha q |x|.
\]
Since \(\alpha<1\) is fixed, applying a Chernoff bound to \(T(J)/|x|\) (a sum of independent random variables in \([0, 1]\) with mean at most \(\alpha q\)) gives, for all sufficiently large \(q\),
\[
    \Pr\big[T(J) \ge (q-3)|x|\big] \le e^{-\Omega(q)}.
\]
If \(Q = q\), then \(I\) is one of the at most \(q\) intervals of \(q\) blocks that contain the block of \(c(x)\), and \(T(I) \ge (q-3)|x|\). Taking a union bound over these intervals,
\[
    \Pr[Q=q]\le q e^{-\Omega(q)} \le e^{-\Omega(q)},
\]
and therefore both \(\E[Q]=O(1)\) and \(\E[Q^2] = O(1)\).
Finally, \(W_x \le |x| Q\), so \(\E[W_x] = O(|x|)\) and \(\E[W_x^2] = O(|x|^2)\).
\end{proof}

\thmSimpleCatalan*
\begin{proof}
Each cell stores a $3$-bit tag, a home bit, and a stop bit. Consider a positive query for an item $x \in S$, and define $Y$ and $\HT_Y$ as above. By \Cref{lem:simple-domination}, the query has the same scan cost in $\HT$ as in $\HT_Y$. In $\HT_Y$, the query scans from $c(x)$ through the last word of $x$, all of which lies in the run containing $c(x)$, so its scan cost is at most $W_x$. Thus, by \Cref{lem:simple-run-length}, there is a constant $C$ (depending only on the load factor) such that a positive query for any item $x \in S$ has expected scan cost at most $C|x|$. By \Cref{lem:stop-marks}, it follows that any negative query also has expected scan cost at most $1 + C = O(1)$.
\end{proof}

\begin{corollary}[Simple Catalan Hashing in External Memory]
\label{cor:intro-simple-em}
Under the working-set assumption of \Cref{sec:preliminaries}, the hash table of \Cref{thm:intro-simple} is cache-oblivious and answers positive queries for any item $x \in S$ in expected $1 + O(|x|/B)$ I/Os and negative queries in expected $1 + O(1/B)$ I/Os.
\end{corollary}

It is worth remarking on why the analysis in this section does \emph{not} extend to get an $O(\epsilon^{-1}|x|)$ bound at high load factors $1 - \epsilon$. Recall that, in a standard linear probing hash table (with unit-sized items), the expected length of the run containing a given cell is $\Theta(\epsilon^{-2})$. Now, consider a table in which all but one item has size $1$, and one item $x$ has size $2$. The item $x$ has lower priority than any other item, and will therefore go at the end of the run formed by elements of size $1$. But this forces a scan length of $\Omega(\epsilon^{-2}|x|)$, which has a quadratic dependence on the load factor. We will see in \Cref{sec:fair-catalan} that this can be avoided by using a more sophisticated placement rule.

\section{Write Optimization via Erasure Codes (Dynamic, Low Load)}
\label{sec:erasure-oracle}

Next we turn to the question of how to support insertions in the Catalan hash table. In this section and the next we focus on the insertion-only regime and defer deletions to \Cref{sec:dynamic-high-load,app:deletions}. Additionally, for this section only, we streamline our presentation by considering only positive queries (not negative ones). 

\paragraph{Why Naive Dynamization Fails.}
At first glance, it might seem simple to dynamize simple Catalan hashing: whenever a new item $x$ is inserted, we could simply update the table so that its layout matches the static simple Catalan table on the current set of items.

The most straightforward way to do so would be when inserting an item \(x\), to insert it and evict all cells with words from lower priority items that block its placement. These evicted words are then shifted and placed, possibly triggering further evictions of even lower priority items, and so on.
However, this naive dynamization incurs substantial overhead.
To see why, suppose the table currently stores a long item $y$ of length $|y| = L \gg B$ in cells $1, 2, \ldots, L$. Now suppose we insert a tiny item $x$ of length $|x| = 1$ with hash $c(x) = 1$. Because $|x| < |y|$, $x$ has higher priority than $y$ and must be placed in cell $1$. This forces all \(L\) words of \(y\) to shift, incurring a scan cost of \(L + 1\) and an I/O cost of \(1+ \Omega(L)/B\).

One could try to improve this cost by allowing the words of \(y\) to be stored out of order. Then in the prior example, the first word of \(y\) could be evicted to the next available cell, while the other words of $y$ stay in place. However, this doesn't change the scan cost, since the evicted word of \(y\) has to be placed after the other \(L\) words. Similarly, in the external-memory model this incurs a random access, which in general adds an I/O.

In the external-memory literature, this phenomenon in which a small logical update triggers a much larger physical reorganization of memory is known as \defn{write amplification}. The standard way to combat write amplification in external-memory search trees (such as $B^\eps$-trees~\cite{BrodalFagerberg2003LowerBounds}, LSM-trees~\cite{ONeilChengGawlickONeil1996LSM}, and cache-oblivious streaming B-trees~\cite{BenderEtAl2007StreamingBTrees}) is to introduce \emph{buffers} at the nodes of a tree, which collect small updates until enough have accumulated to amortize the cost of moving them down a level.

In a linear-probing hash table, however, there is no tree hierarchy, and we cannot redirect insertions to an external buffer without breaking the core guarantee that queries execute in a single contiguous scan of $1 + O(|x|/B)$ I/Os. How, then, can we perform write optimization inside a flat, inline array?

\subsection{Using Large Items as Write Buffers With Erasure Coding}
\label{subsec:erasure-idea}

Our key insight is that we can use \defn{erasure codes} to turn the large items themselves into \emph{in-place write buffers} for smaller items.

Specifically, we use a standard maximum distance separable (MDS) erasure code (such as a Reed-Solomon code) as a black-box primitive~\cite{ReedSolomon60}. 
The MDS code generates, for each item \(x\), \(2|x|\) codewords \(\tilde{x} = \{\tilde{x}_1,\ldots,\tilde{x}_{2|x|}\}\) with the property that we can decode \(x\) from any \(|x|\) of the \(\tilde{x}_i\).
We then store \(\tilde{x}\) as the representation of \(x\) in the table, and allow items of higher priority to simply overwrite the codewords.
The MDS property ensures that as long as at most \(|x|\) words are overwritten, queries can still decode \(x\). The basic idea is that, until $|x|$ words are overwritten, the item $x$ does not need to be moved at all; and once $|x|$ words have been overwritten, we can afford the cost of ``reinflating'' $x$, that is, of reinserting $x$ from scratch to once again have a codeword of length $2|x|$.

One issue that proves (surprisingly!) challenging is how to detect when more than $|x|$ words have been overwritten. A natural approach would be to keep a counter (perhaps at the start of $x$) keeping track of how many codewords have been rewritten. But if a small item overwrites a few codewords in the middle of $x$, it cannot afford to update such a counter, as that would dramatically increase the current operation's scan cost. To separate this issue (determining when more than $|x|$ words have been overwritten) from the other algorithmic issues involved, we assume in this section that we have an oracle that simply tells us when more than half of the codewords for a given element $x$ have been overwritten (we will describe this oracle in more detail shortly). We then devote Section \ref{sec:backup-table} to removing the oracle assumption.

Note that the erasure-coding approach fundamentally requires operating at a low load factor, since we are effectively doubling the size of every item.
Moreover, the metadata scheme we use will add an additional \(O(1)\) words to each cell to enable parsers to distinguish words of different items.
Additionally, in \Cref{sec:backup-table}, we will further add an additional constant-factor in space overhead due to the backup-table technique that we introduce. Thus, the constructions in this section and the next are fundamentally limited to low constant load factors. We will, however, be able to get analogous results at high load factors using the additional techniques introduced in \Cref{sec:fair-catalan,sec:dynamic-high-load}.

\subsection{Warmup: Using an Erasure Oracle}
\label{subsec:oracle-warmup}

The rest of this section will be a warmup that uses an erasure oracle to sidestep the challenge of dealing with too many overwritten codewords.
The \defn{erasure oracle} notifies the algorithm when the number of surviving codewords of a stored item \(x\) drops to \(|x|\) and provides a copy of the original item. 
The algorithm incurs no cost from the oracle.
\Cref{sec:backup-table} removes the oracle assumption. As noted earlier, we focus here on insertions and positive queries only (no deletions or negative queries).

\paragraph{Metadata.}
Overwriting cells in place can break the parser for the static Catalan representation (\Cref{sec:catalan-rep}).
A higher-priority insertion $x$ can overwrite \emph{any} cell of a lower-priority item $y$, including $y$'s first (\copen{}) or last (\cclose{}) cell, which would orphan $y$'s remaining \cmiddle{} cells.
Moreover, MDS erasure decoding requires knowing \emph{which codeword} each surviving cell holds.

We therefore use a different metadata scheme that encodes for each cell containing a codeword \(\tilde{y}_j\) of \(y\):
\begin{itemize}
    \item \(y\)'s item length \(|y|\),
    \item the codeword index \(j \in \{1,\ldots,2|y|\}\), and
    \item a unique owner identifier, \(\ID(y)\).
\end{itemize}
For this warmup, we take \(\ID(y)\) to simply be the number of insertions performed before \(y\), which is less than \(m\).
Thus the metadata for each cell is \(O(1)\) words. 
The ID is used to distinguish the items corresponding to different cells, so we cannot simply use the canonical cell (or a truncated hash) because then collisions would make the resulting cells indistinguishable.
We also change the tie-breaking mechanism for items of the same length to use the ID rather than the hash (as in \Cref{sec:simple-catalan}).
This is because we want to be able to determine the relative priority of items directly from the metadata.

Insertions and queries use a simple parser that separates words of items using their IDs and tracks them with a stack.

\paragraph{Insertions and Reinflations.}
To perform $\ins(x)$, first encode the item $x$ into \(\tilde{x}\). Let \(A_{\succ x}\) denote the set of cells holding codewords of items \(y \succ x\). We say a cell is \defn{available} to \(x\) if it is not in \(A_{\succ x}\), that is, if it is empty or holds a codeword of \(x\) or of a lower-priority item.
Starting at \(c(x)\), scan forward and write \(\tilde{x}_1,\ldots,\tilde{x}_{2|x|}\) into the first \(2|x|\) cells available to \(x\). 
Whenever a cell containing a word of a lower-priority item \(y\) is overwritten, the algorithm asks the oracle if only \(|y|\) codewords of \(\tilde{y}\) are still present.
If so, this triggers an immediate \defn{reinflation} of \(y\).

To perform $\reinflate(y)$, we simply run $\ins(y)$ again using the copy of $\tilde{y}$ obtained from the erasure oracle, keeping the same identifier \(\ID(y)\) (so \(y\)'s priority does not change). Since cells holding codewords of \(y\) are available to \(y\), we will see in \Cref{lem:prefix-occupancy} that this overwrites all remaining codewords of \(y\) from before the reinflation.

\paragraph{Queries.}
As noted earlier, in this section, we only describe positive queries. 
The function $\query(k)$ starts scanning at \(c(k)\), parsing cells and collecting codewords for the items that it reads.
Whenever it obtains \(|y|\) codewords for an item \(y\), it decodes it and compares \(k\) with \(y(k)\) until it finds a matching item.

\subsection{Analysis}
In this section we prove results in the slightly more general case that \(\HT\) operates as above, but may decide to perform \(\reinflate(x)\) at any time when there are at most \(|x|\) surviving codewords (this will be helpful for extending the analysis in later sections).
We will also slightly abuse notation and define the query cost of \(x\) to be the scan cost of reading to the cell that contained the last codeword of \(x\) when $x$ was most recently inserted or reinflated. 
Using the erasure oracle guarantees that \(\reinflate(x)\) is performed any time there are exactly \(|x|\) surviving codewords, and positive queries always terminate at or before the above cell.

Recall that a cell is said to be available to \(x\) if it is not in \(A_{\succ x}\) (i.e., it does not contain a codeword of an item $y$ with higher priority than $x$). Furthermore, define the priority of a cell to be the priority of the item whose word it holds, and \(0\) if the cell is empty. The following two lemmas state basic structural invariants in terms of these definitions.

When interpreting the following lemma, note that the algorithm never clears a cell; it only writes codewords. We will see in \Cref{lem:prefix-occupancy} that a reinflation of \(x\) overwrites all remaining codewords of \(x\) anyway, but by convention, any old codeword that a reinflation did not overwrite would simply remain in place.

\begin{lemma}\label{lem:cell-monotonicity}
     The priority of each cell never decreases over time.   
\end{lemma}

\begin{proof}
    Since cells are never cleared, a cell changes only when an item \(w\) writes to it. The cell is then available to \(w\), so its priority was no higher than that of \(w\). (Equality occurs when a reinflation of \(w\) overwrites an old codeword of \(w\).)
\end{proof}

\begin{lemma}\label{lem:prefix-occupancy}
    At a fixed point in time, if there are \(s\) surviving codewords of \(x\) in \(\HT\), then they occupy the first \(s\) cells available to \(x\) at or after \(c(x)\). In particular, a reinflation of \(x\) overwrites all remaining codewords of \(x\), so the table never holds stale codewords.
\end{lemma}

\begin{proof}
    Consider the most recent insertion or reinflation of \(x\), and let \(r\) be the cell where it writes \(\tilde{x}_{2|x|}\). By construction, it writes the codewords of \(x\) into the first \(2|x|\) cells available to \(x\) at or after \(c(x)\), so afterward every cell in \([c(x), r]\) is either in \(A_{\succ x}\) or holds a codeword of \(x\). 
    
    Now consider what has happened to the cells in $[c(x), r]$ since $x$ was last inserted or reinflated. A cell containing a codeword for $x$ can only be overwritten by an item to which it is available. Thus an item of lower priority than \(x\) cannot overwrite any of $x$'s codewords, and an item of higher priority than \(x\) that overwrites a codeword of \(x\) moves that cell into \(A_{\succ x}\). Once a cell enters $A_{\succ x}$, it stays there by \Cref{lem:cell-monotonicity}. Thus, at the current time, the cells in \([c(x), r] \setminus A_{\succ x}\) are exactly those containing surviving codewords for $x$.

    Finally, we must argue that when the next reinflation occurs, all of $x$'s surviving codewords get overwritten (no stale codewords are left behind). We have already shown, however, that $x$'s surviving codewords from last time it was inserted or reinflated are among the first $2|x|$ cells available to $x$; since the reinflation overwrites all of those cells, it must overwrite all of $x$'s surviving codewords.
\end{proof}

Next we turn to analyzing queries. We argue that we can directly bound the scan cost for a query by what the scan cost would be in a \emph{static table} if every item $x$ was resized to have size $2|x|$. To formalize, this, we introduce the following notation. Let \(S\) be the set of items present at the end of the operation sequence; since there are no deletions, every item that is ever inserted is in \(S\).
For an item $x \in S$, let \(Y = \{y \in S \mid |y| \le |x|\}\), which includes \(x\) and every item of higher priority than \(x\).
Let \(\widetilde{Y} = \{\tilde{y} \mid y \in Y\}\), where we view \(\tilde{y}\) as an item of length \(2|y|\) with hash \(h(y)\).
Let \(\HT_{\widetilde{Y}}\) be a static Catalan hash table on \(\widetilde{Y}\), and let \(L_{\tilde{x}}\) be the length of the run in \(\HT_{\widetilde{Y}}\) containing \(c(x)\).
Since \(\load(\widetilde{Y}) \le 2\load(S)\), if \(\load(S) \le 1/2 - \Omega(1)\) then \Cref{lem:simple-run-length} (applied to \(\tilde{x}\)) gives \(\E[L_{\tilde{x}}] = O(|x|)\) and \(\E[L_{\tilde{x}}^2] = O(|x|^2)\).

\begin{lemma}\label{lem:dynamic-static-domination}
    Let \(x \in S\). The query scan cost of \(x\), and the scan cost of an insertion or reinflation of \(x\) (excluding the reinflations that it triggers), are at most \(L_{\tilde{x}}\).
\end{lemma}

\begin{proof}
    Consider an insertion or reinflation of \(x\), and let \(r\) be the cell where it writes its last codeword. To prove the lemma, it suffices to show that \([c(x), r]\) lies within the run of \(\HT_{\widetilde{Y}}\) containing \(c(x)\).

    At the moment the insertion or reinflation finishes, every cell of \([c(x), r]\) is either in \(A_{\succ x}\) or holds a codeword of \(x\). For each item \(y \succ x\), let \(\overline{y}\) be the item formed from the surviving codewords of \(y\) at that moment, with hash \(h(y)\), and let \(\overline{Y}\) be the set of these items together with \(\tilde{x}\). By \Cref{lem:prefix-occupancy}, each item of \(\overline{Y}\) occupies the first cells at or after its canonical cell that are not occupied by items of higher priority. So the cells of \(A_{\succ x}\), together with the cells of \(x\), are exactly the occupied cells of the bottom-up placement of \(\overline{Y}\) (as in \Cref{sec:simple-catalan}) in the priority order \(\succ\).

    In any such placement, a cell \(c\) is empty if and only if, for every \(j\), the items hashing into \([c-j, c]\) have total length at most \(j\); in particular, the set of occupied cells does not depend on the priority order. Each item of \(\overline{Y}\) is no longer than the corresponding item of \(\widetilde{Y}\) and has the same hash, since every item \(y \succ x\) satisfies \(|y| \le |x|\). So every cell occupied in the placement of \(\overline{Y}\) is also occupied in \(\HT_{\widetilde{Y}}\). In particular, every cell of \([c(x), r]\) is occupied in \(\HT_{\widetilde{Y}}\), so \([c(x), r]\) lies within the run containing \(c(x)\).
\end{proof}

\paragraph{Amortizing Reinflations.}
Finally, we now give a simple argument that bounds the total cost of all reinflations of a given item $x$.
We also prove a second-moment bound on the number of reinflations that we will be helpful later in the paper (in the the proof of \Cref{thm:intro-dynamic-low}).

\begin{lemma}\label{lem:dynamic-reinflations}
Suppose \(\load(S) \le 1/2 - \Omega(1)\).
For an element $x$, let $R_x$ denote the number of times $x$ is reinflated over the entire input sequence of $\HT$ and 
$F_x$ be the query scan cost in $\HT$ immediately after the last reinflation of $x$.

Then 
\[ \E[R_x^2] = O(1) \qquad \textrm{and} \qquad \E[R_xF_x] = O(|x|).\]
\end{lemma}

\begin{proof}
    Let $F_x(t)$ denote the query scan cost of $x$ after the $t$-th reinflation.
    Since \(|x|\) cells belonging to \(x\) must be overwritten between reinflations, \Cref{lem:prefix-occupancy,lem:cell-monotonicity} imply \(F_x(t+1) - F_x(t) \ge |x|\) for all \(t \leq R_x\).
    Therefore \(R_x \le F_x/|x|\).

    By \Cref{lem:dynamic-static-domination}, \(F_x \leq L_{\widetilde{x}}\).
    Therefore, by \Cref{lem:simple-run-length},
    \[
        \E[R_x^2] \le \E[F_x^2]/|x|^2 \le \E[L_{\tilde{x}}^2]/|x|^2 = O(1),
    \]
    and
    \[
        \E[R_xF_x] \le \E[F_x^2]/|x| \le \E[L_{\tilde{x}}^2]/|x| = O(|x|).
    \]
\end{proof}

Since $R_x \cdot F_x$ in the lemma above is a bound on the total scan cost of all reinflations of $x$, we can now complete the analysis of the data structure:

\begin{theorem}[Dynamic Catalan Hashing in the Erasure Oracle Model]
\label{thm:oracle-model}
\label{lem:oracle-reinflation-amortized}
In the erasure oracle model at load factor $\alpha \leq 1/2-\Omega(1)$, there is an insertion-only dynamic Catalan hash table in which a positive query for an item $x$ has expected scan cost $O(|x|)$, and an insertion of $x$ has amortized expected scan cost $O(|x|)$.
\end{theorem}

\begin{proof}
By \Cref{lem:dynamic-static-domination} and the bound \(\E[L_{\tilde{x}}] = O(|x|)\), a positive query for \(x\) scans \(O(|x|)\) cells in expectation. We now bound the expected amortized cost of reinflations.
Let \(R_x\), \(F_x(t)\), and \(F_x\) be as in the preceding lemma.
The algorithm performs \(R_x\) reinflations, each of which has scan cost at most \(F_x\).
Therefore the expected amortized cost of reinflations of 
\(x\) is \(\E[R_xF_x] = O(|x|)\) by \Cref{lem:dynamic-reinflations}. We charge the cost of these reinflations to the initial insertion of $x$.
Ignoring the cost of reinflations that an insertion triggers, insertions have expected scan cost \(O(|x|)\) by \Cref{lem:dynamic-static-domination}. Thus the overall amortized expected scan cost of an insertion (including the costs of reinflations) is \(O(|x|)\).
\end{proof}

\section{Removing the Oracle via Lazy Reinflation from a Backup Table}
\label{sec:backup-table}

In \Cref{sec:erasure-oracle}, we constructed a dynamic Catalan hash table that used an erasure oracle to prevent items from being overwritten too much.
In this section, we show how to remove the oracle. Rather than trying to simulate the oracle directly (which is difficult to do without increasing scan cost, as explained in \Cref{subsec:erasure-idea}
), we remove the oracle by taking a different approach.

When we insert items, we insert them both into the actual table, and into a second backup table that is interleaved with the actual table. The backup table is less efficient than the full table (incurring roughly $\log^{1+\delta} s$ factor overhead on all operations), and is used by queries \emph{only} when query cannot recover enough codewords in the main table to reconstruct the element $x$ it is looking for. When this happens, we recover $x$ from the backup table, and reinflate $x$. We call this a \emph{lazy reinflation}, since the reinflation is only performed when necessary when a query actually fails to be able to reconstruct $x$. Critically, we show that we can charge the cost of lazy reinflations to the original insertion of $x$; this allows us to charge the inefficient-ness of the backup table (which is inefficient for \emph{both} insertions and queries) to insertions only, without affecting the amortized cost of queries. 

To make the above construction work with negative queries we also introduce another interleaved table that stores fingerprints of present items. This enables queries to determine membership (whether the element queried is present) efficiently, and therefore allows negative queries to (most likely) avoid having to go into the backup table at all.

\subsection{High-Level Construction: How to use the Membership and Backup Tables}\label{sec:backup-reduction}

We now formalize the high-level construction described above. For now, we treat the backup table as a black box, and focus on how to use it within the larger data structure. 

Our starting point is the dynamic Catalan hash table \(\HT\) from \Cref{sec:erasure-oracle} with the same priority order, metadata, and insertion and reinflation procedures as defined in that section. (Below we do change the unique IDs used to index items, but they function in the same way.)

We add two secondary tables interleaved evenly with \(\HT\): a \defn{backup table} \(\backup\), which stores full items (that cannot be overwritten), and a \defn{membership table} \(\member\), which stores compact fingerprints for keys that are present.
All three tables store items using the same hash function \(h\), so that for each key \(k\), \(c(k)\) refers to the three adjacent cells with one cell from each table.
We abuse notation and refer to this triple of cells as \(c(k)\).

Interleaving the three tables is key to maintaining the property that all data is stored in a shared physical array, thus avoiding a blowup in scan costs or I/Os from the addition of these auxiliary tables. Operations will scan forward starting from \(c(k)\), parsing cells from each table, which we think of as running concurrent operations on all three tables (some of which may be be no-ops).

We will assume throughout this section that the load factor is at most a small positive constant of our choice. We refer to the combined interleaved table as \(\CHT(\HT, \member, \backup)\); we will frequently write simply $\CHT$ when the component tables are clear from context.

\paragraph{The Membership Table and Negative Queries.}
The membership table \(\member\) is a standard unordered fixed-length linear-probing hash table.
For each item \(y\in S\), \(\member\) stores a triple \((\phi(k(y)), F_y, |y|)\), where \(\phi(k(y))\) is a fingerprint, and \(F_y\) is the index of the last cell occupied by a codeword of \(y\) at the time of its insertion or last reinflation.
The fingerprints are computed with an independent hash function \(\phi\) and are $C\log m$ bits in length, for a sufficiently large constant $C$.

A query for a key \(k\) begins by scanning \(\member\) from \(c(k)\) to the first empty cell of \(\member\), collecting the entries whose fingerprint is \(\phi(k)\). If there are none, the query returns \textsc{Not Found}. Otherwise, it runs the positive-query procedure described below for each matching entry. Since the tables are interleaved, scanning \(j\) cells of \(\member\) has scan cost \(O(j)\) in \(\CHT\).
The stored length tells the query how many codeword symbols suffice to decode a candidate; we will also use it to support deletions in \Cref{app:deletions}.

We also redefine in this section the unique ID of each item \(x\in S\) to be the index of its position in \(\member\).
Unordered (and insertion-only) linear-probing hash tables are stable (items do not move after insertion), so the IDs do not change over time.  As we will discuss shortly, this re-definition of ID is helpful to positive queries for determining which codewords in the main table $\mathcal{H}$ correspond to the item $x$ being queried. 

The following lemma shows that negative queries are efficient in this construction.
\begin{lemma}\label{lem:dynamic-negative}
  A negative query to \(\CHT\) has expected scan cost \(O(1)\).
\end{lemma}

\begin{proof}
    Let \(k\) be a key not in the table. Since \(\member\) holds one entry per item of \(S\), it is a unit-size linear-probing table at load factor at most \(\load(S) < 1 - \Omega(1)\). So the run of \(\member\) containing \(c(k)\), and hence the scan of \(\member\) from \(c(k)\) to the first empty cell, has expected length \(O(1)\) by \Cref{lem:simple-run-length} (applied with unit-length items, with \(k\) added as one of them). The query returns \textsc{Not Found} at the end of this scan unless some item \(y \in S\) has \(\phi(k(y)) = \phi(k)\). Each item collides with \(k\) with probability \(m^{-C}\), so by a union bound a collision occurs with probability at most \(m^{1-C}\). Every operation has scan cost at most polynomial in \(m\), so for a sufficiently large constant \(C\), collisions contribute \(O(1)\) to the expected scan cost.
\end{proof}

\paragraph{The Backup Table and Lazy Reinflation.}
The backup table \(\backup\) stores a second copy of each item \(x\in S\) that cannot be overwritten, together with its ID.
It will have potentially worse performance guarantees, incurring expected scan cost \(O(|x|\cdot q(|x|))\) for some function \(q\) for both insertions and queries.
However, in \(\CHT\), we will only pay this cost on insertions.
For now, we'll treat \(\backup\) as a black box and defer its construction to \Cref{sec:backup-construction}.

When an item \(x\) is inserted, we insert \(\tilde{x}\) into \(\HT\), \(x\) into \(\backup\), and \((\phi(x),F_x,|x|)\) into \(\member\).
When a subsequent insertion overwrites cells of \(\tilde{x}\), we do nothing.
We then lazily reinflate during queries as follows.
A query for \(x\) finds \(\phi(x)\) and \(F_x\) in \(\member\), and computes \(\ID(x)\).
If the query finds \(|x|\) codewords in \(\CHT\) while scanning from \(c(x)\) to \(F_x\), then it can decode \(x\) and return \textsc{Found} together with \(v(x)\).
Otherwise, this means that more than \(|x|\) codewords have been overwritten, so \(x\) requires reinflation.
The query finds \(x\) in \(\backup\), then performs \(\reinflate(x)\) in \(\CHT\).
Then it updates \(F_x\) in \(\member\).

\paragraph{Handling False Positives in \(\member\).}
For a matching membership entry at position $p$, the query only collects codewords from the primary table with owner identifier $p$.
If too few survive, it searches the backup table for the item with identifier $p$.
In either case, it compares the recovered key with $k$ before returning the item or reinflating it; a mismatch rejects this entry.
A backup search stops once it has read the complete candidate item, allowing at most two laps of the table so that a chunk crossing the starting point can be read in full.
For the true candidate, the search starts at its canonical cell and has the usual positive-query scan bound.
A false candidate may require a full-table scan, but this occurs only on a fingerprint collision.
This happens rarely enough that it does not affect the expected scan bounds (as in the proof of \Cref{lem:dynamic-negative}, we can even afford to scan the entire table in the event of a fingerprint collision, which happens with probability \(m^{-(C-1)}\)).

\subsection{Amortizing Reinflations Against Insertions}
When a query for \(x\) triggers a reinflation, the query must retrieve \(x\) from the backup table \(\backup\) incurring a scan cost of \(O(|x|\cdot q(|x|))\), then execute \(\reinflate(x)\) in \(\CHT\).
At first glance, this appears to make queries as expensive as backup-table queries.

The key observation is that we can amortize the cost of these reinflations against the insertions.
Then in this amortized sense, a query for \(x\) still incurs a scan cost of \(O(|x|)\) while \(x\)'s initial insertion pays for the backup-table queries and reinflations.
The reason this works is that once \(x\) has been reinflated, subsequent queries for \(x\) will be answered from the main table \(\CHT\) rather than requiring the expensive reinflation.
Only when sufficiently many higher priority items again over-overwritten \(x\)'s codewords in \(\CHT\) will a query for \(x\) need to reinflate again.

\begin{lemma}[Backup-Table Reduction]
\label{lem:backup-reduction}
    Let \(\alpha_0 \le 1/2 - \Omega(1)\) be a positive constant, and suppose that \(\load(S) \le \alpha_0\), where the load factor is measured relative to the \(m\) cells of \(\HT\).
    Let \(\backup\) be a dynamic hash table with \(m\) cells that supports insertions and positive queries (queries for keys in \(S\)).
    For \(x\in S\), let \(\Lambda_x\) be the random variable of the maximum scan cost of any insertion or query of $x$ in \(\backup\) over the entire operation sequence.
    Suppose \(\E[\Lambda_x] = O(|x|\cdot q(|x|))\) and \(\E[(\Lambda_x)^2] = O((|x| \cdot q(|x|))^2)\) for all \(x \in S\).

    Let \(\CHT = \CHT(\HT,\member,\backup)\). 
    Then 
    \begin{itemize}
    \item If \(x \in S\), \(\query(k(x))\) in \(\CHT\) has amortized expected scan cost \(O(|x|)\);
    \item If \(k \not\in S\), \(\query(k)\) in \(\CHT\) has expected scan cost \(O(1)\); and
    \item \(\ins(x)\) has amortized expected scan cost \(O(|x|\cdot q(|x|))\).
    \end{itemize}
\end{lemma}

\begin{proof}
    The negative-query bound is \Cref{lem:dynamic-negative}.
    Since \(\load(S) \le \alpha_0 \le 1/2 - \Omega(1)\), \Cref{lem:dynamic-static-domination} and the bound \(\E[L_{\tilde{x}}] = O(|x|)\) from \Cref{sec:erasure-oracle} apply to \(\HT\).
    Therefore, the scan cost of a positive query that doesn't trigger a reinflation is \(O(|x|)\).

    Let \(R_x\) be the number of reinflations and \(F_x\) be the query cost of \(x\) in \(\HT\) after the last reinflation.
    The total amortized cost of reinflation is at most \(R_x(\Lambda_x + F_x)\).
    Taking expectations and applying Cauchy-Schwarz, we have
    \begin{equation}\label{eqn:reinflation-cs}
    \E[R_x(\Lambda_x + F_x)] \le \sqrt{\E[R_x^2]\E[\Lambda_x^2]} + \E[R_xF_x].
    \end{equation}
    \(\HT\) satisfies the hypotheses of \Cref{lem:dynamic-reinflations}, so
    \(\E[R_x^2] = O(1)\) and \(\E[R_xF_x] = O(|x|)\).
    Therefore also using the hypothesis on \(\Lambda_x\), the RHS of \Cref{eqn:reinflation-cs} is \(O(|x| \cdot q(|x|))\). We charge this cost to the insertion of \(x\), which itself has expected scan cost \(O(|x|)\) in \(\HT\), \(O(1)\) in \(\member\), and \(\E[\Lambda_x] = O(|x| \cdot q(|x|))\) in \(\backup\).
\end{proof}

\subsection{Constructing the Density-Regulated Backup Table}
\label{sec:backup-construction}
\label{subsec:putting-low-dynamic-together}

With \Cref{lem:backup-reduction} in place, the only remaining piece of \Cref{thm:intro-dynamic-low} is the backup table itself. The goal of this subsection is to construct a backup table $\backup$, which we call the \defn{density-regulated backup table}, that satisfies the hypotheses of \Cref{lem:backup-reduction} with $q(s) = \log^{1+\delta} s$. That is, $\backup$ needs to support only insertions and positive queries, and for each item $x$ of length $s$, the insertion of $x$ and every query for $x$ should incur scan cost roughly $O(s \log^{1+\delta} s)$, in the sense that the maximum $\Lambda_x$ of these scan costs satisfies $\E[\Lambda_x^2] = O((s \log^{1+\delta} s)^2)$. (Throughout this subsection, $\log s$ denotes $\max\{1, \log_2 s\}$.)

What makes this task easier than the one we faced in the primary table is that $\backup$ is allowed to lose a $\log^{1+\delta} s$ factor on \emph{both} insertions and queries. This, it turns out, allows for a completely different (and somewhat simpler) construction than those in previous sections, one in which items never overwrite or move one another.

\paragraph{The Basic Idea.}
The basic idea is to separate items by size, and to divide $\backup$ among the resulting size classes in proportion to their volumes. We partition the items into \defn{size classes} $C_i = \{x \mid 2^i \le |x| < 2^{i+1}\}$, and we give each class a few strided regions of $\backup$, which we call \defn{subarrays}: a subarray of \defn{density} $2^{-k}$ consists of every $2^k$-th cell of $\backup$. When a class $C_i$ receives its first item, it is given a subarray of density roughly $1/(i+1)^{1+\delta}$, and whenever the class fills half of its newest subarray, it is given a new subarray of twice the density. Since $\sum_i 1/(i+1)^{1+\delta}$ converges, the first subarrays of all of the classes together take up only a small constant fraction of $\backup$. Since additional subarrays are allocated on a per-demand basis (only when a size class has already received a large volume of elements), we will be able to ensure that the total density of all subarrays ever allocated is less than $1$. This will be the critical point that makes the algorithm possible.

Within each subarray, a class stores its items using standard unit-size linear probing. Since the items of $C_i$ have the same length up to a factor of $2$, we can divide the subarray into fixed-size \defn{chunks} of $2^{i+2}$ consecutive subarray cells, each of which holds at most one item. Note that, even though the item is stored contiguously \emph{within the subarray}, it is not stored physically contiguously: a chunk in a subarray of density $1/D$ is physically spread across $\Theta(D 2^i)$ cells, which inflates the scan cost by a factor of $\Theta(D)$. This is the main thing we need to be careful about when bounding scan cost. 

For the first subarray of class $C_i$, the stride (i.e., the physical distance between consecutive cells of the subarray) is roughly $(i+1)^{1+\delta} = O(\log^{1+\delta} s)$, and the later subarrays of the class are only denser, so this ends up being the source of the $\log^{1+\delta} s$ factor when analyzing the scan costs of both insertions and queries.

In the rest of the subsection, we fill in the details for the data structure described above. 

\paragraph{Allocating Subarrays.}
We begin by defining the strided subarrays in which classes store their items, and by defining the simple allocator that we use to hand these strided subarrays out to different classes. 

Let \(m_{\backup}\) denote the number of cells of \(\backup\); with the interleaving above, \(m_{\backup} = m\).
Throughout this subsection, we assume that \(\Vol(S) \le m_{\backup}/64\), that is, \(\load(S) \le 1/64\).

For $0 \le k \le \log m_{\backup}$ and $0 \le a < 2^k$, we define  subarray $\backup_{k,a}$ to consist of the $m_{\backup}/2^k$ cells $a, a + 2^k, a + 2 \cdot 2^k, \ldots$ of $\backup$, in this order. Its density is $2^{-k}$. The subarrays form a complete binary tree: the two children $\backup_{k+1, a}$ and $\backup_{k+1, a + 2^k}$ of $\backup_{k,a}$ partition it, so two subarrays are either disjoint or nested. 

We allocate subarrays with the following simple allocator (essentially a buddy allocator \cite{Knowlton65,Knuth97}). To allocate a subarray of density $2^{-k}$, let $j \le k$ be the largest depth at which some free subarray exists, and let $Q$ be any free subarray at depth $j$. We allocate one of the depth-$k$ subarrays contained in $Q$, and the rest of $Q$, which consists of one subarray at each depth $j + 1, \ldots, k$, remains free. The next lemma shows that this allocator succeeds so long as the total allocated density stays at most $1$.

\begin{lemma}
\label{lem:buddy}
An allocation of density $2^{-k}$ succeeds whenever the total density of all allocated subarrays, including the new one, is at most $1$.
\end{lemma}
\begin{proof}
The key invariant is that, at any moment, there is at most one free subarray at each depth. To see that allocations preserve this, consider an allocation of density $2^{-k}$ that splits a free subarray at depth $j$. The allocation creates one new free subarray at each depth $j + 1, \ldots, k$, and there were no free subarrays at these depths beforehand, since otherwise the allocator would have chosen a larger $j$.

Now suppose that an allocation of density $2^{-k}$ fails, meaning that no free subarray has depth at most $k$. By the invariant, there is at most one free subarray at each depth $d > k$, so the free subarrays have total density less than $\sum_{d > k} 2^{-d} = 2^{-k}$. Since the free subarrays cover all of the unallocated cells, the allocated density is more than $1 - 2^{-k}$, so the allocated density plus $2^{-k}$ exceeds $1$.
\end{proof}

\paragraph{Levels and Chunks.}
We now describe how each class $C_i$ stores its items. The class is stored in a sequence of subarrays, which we call \defn{levels}, whose densities double from each level to the next. Within each level, the class is simply a unit-size linear-probing table whose entries are chunks of $2^{i+2}$ cells of $\backup$. Within a given size class $C_i$, new items are only ever inserted into the newest level, which is never more than half full. (The doubling of densities is not needed to bound scan costs; it just keeps the number of levels of each class at $O(\log m_{\backup})$.)

Fix a constant $c_0 > 0$ such that $\sum_{i \ge 0} c_0/(i+1)^{1+\delta} \le 1/8$, and let $d_i$ be the largest power of two that is at most $c_0/(i+1)^{1+\delta}$. Level $\ell$ of class $C_i$ is a subarray of density $D_{i,\ell} = 2^{\ell} D_{i,0}$, where $D_{i,0} = \max\{d_i,\, 2^{i+3}/m_{\backup}\}$ (the second term just ensures that every level has at least two chunks). It is divided into $n_{i,\ell} = D_{i,\ell}\, m_{\backup}/2^{i+2}$ \defn{chunks} of $2^{i+2}$ consecutive cells, and an item stored in a chunk occupies a prefix of it. This prefix starts with a two-word header containing the item's identifier and length. The header and item fit in the chunk, since $|x|+2\le 2^{i+1}+1\le 2^{i+2}$ for $x\in C_i$. Queries can therefore recognize items by identifier as well as by key, without changing their placement or scan bounds.

Let $\gamma(x)$ be the cell of $\backup$ that $x$ hashes to, namely the cell of \(\backup\) in \(c(x)\), which is uniformly random and independent across items. In each level, the \defn{home chunk} of $x$ is the first chunk (cyclically) whose first cell is at or after $\gamma(x)$. Since the first cells of the chunks are evenly spaced, home chunks are uniformly random and independent across items, and $x$'s home chunk in level $\ell$ begins fewer than $2^{i+2}/D_{i,\ell}$ cells after $\gamma(x)$.

With these definitions in place, the operations are implemented as follows.
\begin{itemize}
    \item \textbf{Insert$(x)$}, for $x \in C_i$: if $C_i$ has no levels, or its newest level $L$ already holds $n_{i,L}/2$ items, allocate a new level. Then write $x$ into the first empty chunk of the newest level at or after $x$'s home chunk.
    \item \textbf{Query$(x)$}, for $x \in S$: scan forward from $\gamma(x)$, parsing the items in the chunks that the scan passes through, until finding $x$.
\end{itemize}
The bookkeeping for the allocator and the levels takes $O(\log^2 m_{\backup})$ words of working space, as used in the I/O analysis below.

\paragraph{Analysis.}
It remains to show that the allocator never runs out of space, and to bound the scan costs of insertions and queries. We begin with space. The key observation is that, beyond level $0$, the levels of a class are paid for by the volume of the class: a class only receives level $L \ge 1$ once level $L - 1$ is half full, at which point the volume of the class is proportional to the volume of the level being allocated.

\begin{lemma}[Density Never Exceeds $7/8$]
\label{lem:backup-density}
At all times, the total density of the allocated subarrays is at most $7/8$. Hence, by \Cref{lem:buddy}, every allocation succeeds.
\end{lemma}
\begin{proof}
We bound the density used by the classes in two groups, depending on whether or not a class has advanced beyond level $0$.

First consider the classes whose newest level is level $0$. Such a class $C_i$ uses density $D_{i,0} \le d_i + 2^{i+3}/m_{\backup}$. Since every item has $|x| \le \Vol(S) \le m_{\backup}/64$, every class satisfies $2^i \le m_{\backup}/64$. Summing over all such classes, the total density is therefore at most
\[
    \sum_{i} d_i + \sum_{i \,:\, 2^i \le m_{\backup}/64} \frac{2^{i+3}}{m_{\backup}} \le \frac{1}{8} + \frac{16 \cdot m_{\backup}/64}{m_{\backup}} = \frac{3}{8}.
\]

Next consider a class $C_i$ whose newest level is $L \ge 1$. At the moment when level $L$ was allocated, level $L - 1$ held $n_{i,L-1}/2$ items, each of length at least $2^i$. Thus the class has volume
\[
    \Vol(C_i) \ge n_{i,L-1} 2^{i-1} = D_{i,L-1}\, m_{\backup}/8 = D_{i,L}\, m_{\backup}/16.
\]
On the other hand, since the densities of the levels double from each level to the next, the total density of the levels of $C_i$ is $\sum_{\ell \le L} 2^{\ell} D_{i,0} < 2 D_{i,L}$, which by the above is at most $32 \Vol(C_i)/m_{\backup}$. Summing over all such classes, the total density is at most $32 \Vol(S)/m_{\backup} \le 32 (m_{\backup}/64)/m_{\backup} = 1/2$.

Adding together the two groups, the total density is at most $3/8 + 1/2 = 7/8$.
\end{proof}

Finally, we bound the scan costs. A key observation is that querying $x$ costs no more than inserting $x$ did: $x$ never moves, and a query for $x$ scans forward from $\gamma(x)$ only until it reaches $x$, which is exactly where the insertion of $x$ stopped. So it suffices to bound insertions. An insertion scans its level, which is a unit-size linear-probing table with load factor at most $1/2$, from $x$'s home chunk to the first empty chunk. This is $O(1)$ chunks in expectation, each spanning at most $2^{i+2}/d_i$ cells of $\backup$, for a total of $O(2^i/d_i) = O(s \log^{1+\delta} s)$.

\begin{lemma}[Performance of the Density-Regulated Backup Table]
\label{lem:backup-cost}
Let $x \in C_i$ have length $s = |x|$. There is a random variable $\Lambda_x$ with $\E[\Lambda_x^2] = O((s \log^{1+\delta} s)^2)$ such that the insertion of $x$, and every query for $x$ over the entire operation sequence, scans a contiguous interval of $\backup$ starting at $\gamma(x)$ of length at most $\Lambda_x$. In particular, $\backup$ satisfies the hypotheses of \Cref{lem:backup-reduction} with $q(s) = \log^{1+\delta} s$ and \(\alpha_0 = 1/64\).
\end{lemma}

\begin{proof}
Let $\ell$ be the level into which $x$ is inserted, and let $J_\ell$ be the number of chunks from $x$'s home chunk through the first empty chunk of level $\ell$ at the time of the insertion; $x$ is written into that empty chunk. The home chunk begins fewer than $2^{i+2}/D_{i,\ell}$ cells after $\gamma(x)$, and each chunk spans at most $2^{i+2}/D_{i,\ell}$ cells of $\backup$, so the insertion of $x$ scans at most
\[
    \Lambda_x = (J_\ell + 1)\, \frac{2^{i+2}}{D_{i,\ell}} \le (J_\ell + 1)\, \frac{2^{i+2}}{d_i}
\]
cells starting at $\gamma(x)$. Since $x$ is never moved, and a query for $x$ stops once it has read $x$, every query for $x$ also scans at most $\Lambda_x$ cells.

It remains to show that $\E[J_\ell^2] = O(1)$. View level $\ell$ as a unit-size linear-probing table whose cells are its chunks, in which each item hashes to its home chunk. Which items are placed in level $\ell$ (including whether $x$ is) depends only on the operation sequence, and not on the hashes, since the level that receives an insertion is determined by the number of earlier insertions into $C_i$. The level holds fewer than $n_{i,\ell}/2$ items before $x$ is inserted, so its load factor is at most $1/2$. If $R$ is the length of the run of occupied chunks containing $x$'s home chunk at the time of the insertion (with $R = 0$ if that chunk is empty), then $J_\ell \le 1 + R$. Applying \Cref{lem:simple-run-length} with all items having length $1$ gives $\E[R^2] = O(1)$. (The proof of that lemma uses only two properties, both of which hold here: that every word of a run belongs to an item whose canonical cell lies in the run, and that the items other than $x$ have independent hashes.) Hence $\E[J_\ell^2] = O(1)$, and so $\E[\Lambda_x^2] = O((2^i/d_i)^2)$.

Finally, $2^i \le s$ and $1/d_i \le 2(i+1)^{1+\delta}/c_0 \le 2(2\log s)^{1+\delta}/c_0$, so $2^i/d_i = O(s \log^{1+\delta} s)$. This establishes the bound on $\E[\Lambda_x^2]$, and by Jensen's inequality, $\E[\Lambda_x] \le \E[\Lambda_x^2]^{1/2} = O(s \log^{1+\delta} s)$ as well. Thus the hypotheses of \Cref{lem:backup-reduction} hold.
\end{proof}

Finally, plugging \Cref{lem:backup-cost} into \Cref{lem:backup-reduction} proves \Cref{thm:intro-dynamic-low}, which we restate here for convenience:

\thmDynamicLow*
\begin{proof}
Choose \(m\) so that \(\load(S) \le 1/64\), and apply \Cref{lem:backup-reduction} with \(\alpha_0 = 1/64\) and \(\backup\) instantiated as the density-regulated backup table on \(m_{\backup} = m\) cells. Then \(\Vol(S) \le m_{\backup}/64\), so by \Cref{lem:backup-cost}, the hypotheses of \Cref{lem:backup-reduction} hold with $q(s) = \log^{1+\delta} s$, which gives the claimed bounds. The three interleaved tables use \(3m\) cells in total, so the load factor is a positive constant.
\end{proof}

\begin{corollary}[Dynamic Low-Load Catalan Hashing in External Memory]
\label{cor:intro-dynamic-low-em}
Suppose $M \ge c_M(s_{\max} + B + \log^2 m)$ for a sufficiently large constant $c_M$. The hash table of \Cref{thm:intro-dynamic-low} is cache-oblivious and answers positive queries for any item $x$ in amortized expected $1 + O(|x|/B)$ I/Os, answers negative queries in expected $1 + O(1/B)$ I/Os, and supports insertions of any item $x$ in amortized expected
\[
    1 + O\left(\frac{|x| \log^{1+\delta}|x|}{B}\right)
\]
I/Os.
\end{corollary}

\begin{proof}
We use the working-set assumption of \Cref{sec:preliminaries}, together with the stated cache-size bound. The additional working space is $O(s_{\max} + \log^2 m)$ words: we collect the codeword symbols read during a scan to encode or decode the item there, and retain the allocator's $O(\log^2 m)$ words of global bookkeeping. Thus encoding and decoding incur no I/Os beyond those of the scan. Each operation makes $O(1)$ forward passes over its scan interval, so the scan property converts the scan bounds of \Cref{thm:intro-dynamic-low} into the stated I/O bounds, with the same amortization. The algorithms do not use $M$ or $B$.
\end{proof}

\section{Fair Catalan Hashing (Static, High Load)}
\label{sec:fair-catalan}

Up to this point, we have focused on the low-load-factor regime, where the load factor $\alpha = \load(S)$ is $1 - \Omega(1)$. We now turn to the \defn{high-load-factor regime}, where the table is filled to load factor
\[
    \alpha = 1 - \eps
\]
for a small parameter $\eps \in (0, 1)$, and prove \Cref{thm:arch}.

In classical unit-size linear probing, standard (unordered) placement suffers from primary clustering (where runs of items merge together), resulting in an expected query cost of $\Theta(\eps^{-2})$~\cite{Knuth98}. By contrast, ordered linear probing~\cite{AmbleKnuth74} (more popularly known as Robin Hood hashing~\cite{CelisLarsonMunro1985RobinHood}), which gives priority to items with smaller hash values, reduces the worst-case expected query cost for any item to $\Theta(\eps^{-1})$.

When items have variable lengths, however, simple Catalan hashing gets stuck at $\Theta(\eps^{-2})$. 
To see this, consider an $m$-cell table containing $(1 - \eps)m - 2$ items of length $1$ and a single item $x$ of length $2$. Because simple Catalan hashing prioritizes shorter items over longer ones, every unit-length item has higher priority than $x$. Consequently, when $x$ is placed, it must skip over all unit-length items in its run and place its words into the first two empty cells left behind by the unit-length items. By the standard analysis of unit-length linear probing at load factor $(1 - \eps)$~\cite{Knuth98}, the expected distance from a random cell $c(x)$ to the first empty cell is $\Theta(\eps^{-2})$, so querying $x$ requires scanning $\Theta(\eps^{-2})$ cells in expectation.

At the same time, we cannot simply switch to Robin Hood priority (giving priority to items with smaller hash values), because in the variable-length setting, a single long item $y$ of length $L$ would have higher priority than all items hashing after $h(y)$, forcing short items of length $1$ to scan past all $L$ words of $y$.
Thus, \emph{neither} prioritizing by length (simple Catalan hashing) \emph{nor} prioritizing by hash value (Robin Hood hashing) can achieve a linear $O(\eps^{-1})$ dependence on the inverse slack.

\subsection{Hash-Table Queues: Variable-Length Linear Probing as Preemptive Queuing}
\label{subsec:scheduling-view}

The reason that simple Catalan hashing and Robin Hood hashing don't achieve the query bounds we would like in the high-load case is that they are \emph{unfair}: some jobs unfairly take priority over others.
Robin Hood hashing gives too much priority to large jobs, so small jobs have high expected query cost.
On the other hand, simple Catalan hashing gives too much priority to small jobs, so large jobs have high expected query cost.
This is analogous to fairness problems in queuing theory, where a long job can block short jobs, while strict priority for short jobs can penalize long ones.
We show that there is a surprisingly direct connection from variable-length linear probing to fair queuing and develop \defn{fair Catalan hash tables} based on fair queuing policies.

We can reinterpret variable-length linear probing as a \defn{preemptive single-server scheduling problem}, where we view the hash table $[0, m) \subset \mathbb{R}$ as a continuous time axis.
We define the following:

\begin{description}[labelindent=1em, leftmargin=2.5em, style=sameline]
    \item[Jobs:] Each \defn{job} is an item $x \in S$.
    \item[Arrival times:] A job \(x\)'s \defn{arrival time} \(A_x\) is \(h(x)\).
    \item[Service requirements:] A job \(x\)'s \defn{service requirement} (or size) is \(|x|\).
    \item[Schedule:] A job \(x\)'s schedule is its placement in the table. The words of \(x\) correspond to units of time \(x\) is serviced (with preemption allowed).
    \item[Departure times: ] A job \(x\)'s \defn{departure time} \(D_x\) is the position of its last cell.
    \item[Sojourn times: ] A job \(x\)'s \defn{sojourn time} (or response time) is the query scan cost \(L_x\). Its \defn{stretch} (or slowdown) is \(L_x/|x|\).
\end{description}

We call the resulting queuing model the \defn{hash-table queue}.
In this model, the set of jobs and their sizes are fixed, and the arrival times are independent and uniform on a circle of circumference $m$.
Service in a hash-table queue is interpreted cyclically: since the total service requirement is $\alpha m<m$, the schedule has idle time at which we can unwrap the circle.

For comparison, the classical \defn{M/G/1 queue} has a Poisson stream of arrivals and independent, identically distributed service requirements, independent of the arrival process~\cite{Kleinrock76,HarcholBalter13}.
The notation refers to memoryless (exponentially distributed) interarrival times ($M$), a general distribution of service requirements ($G$), and a single server ($1$).
The server supplies one unit of service per unit of time, and a scheduling policy specifies how that service is allocated among the jobs present.
The load $\rho$ is the arrival rate times the mean service requirement; the stable regime $\rho<1$ is analogous to $\alpha<1$ for hash-table queues.

Interpreting linear probing as a hash-table queue, different linear-probing priority rules correspond directly to different scheduling policies:
\begin{itemize}
    \item \textbf{Ordered linear probing (Robin Hood)} is \textbf{First-Come-First-Served (FCFS)}: Jobs are served in order of arrival time $A_x$, without preemption. FCFS is unfair to small jobs that arrive behind a large job.
    \item \textbf{Simple Catalan hashing} is \textbf{Preemptive Static-Size Priority}: Jobs with smaller original size $|x|$ have strict preemptive priority over jobs with larger original size. This is unfair to large jobs, which get starved until the very end of the busy period (run), incurring $\Theta(\eps^{-2})$ stretch.
\end{itemize}
What we want is a scheduling policy that is \emph{fair} across all job sizes---guaranteeing expected stretch $O(\eps^{-1})$ for every item $x$, regardless of $|x|$---while still producing a table that satisfies the Catalan nesting property (\Cref{def:catalan-property}).

For this we look to fair queues.
We first consider a policy that gives the desired bound for each item's expected stretch, and then a policy based on processor sharing that also bounds the expected maximum stretch over the table.

\subsection{Anti-Robin-Hood Catalan Hashing (ARCH)}
In queuing theory, the simplest policy whose expected slowdown is $O(\eps^{-1})$ independent of job size is \defn{Last-In-First-Out with Preemption (LIFO-PR)}, where newly arriving jobs immediately preempt whatever job is currently in service. In our hash-table setting, LIFO-PR corresponds to what we call \defn{anti-Robin-Hood Catalan hashing (ARCH)}, so named because its priority order is the exact \emph{opposite} of that of Robin Hood hashing. Formally, the ARCH hash table is the static Catalan hash table of \Cref{def:catalan} in which an item $x$ has priority over an item $y$ if $h(x) > h(y)$, where hashes are compared cyclically.

We remark that, in the context of Catalan hashing, it may seem counterintuitive that the \emph{anti-Robin-Hood} order is in some sense ``fairer'' than the Robin Hood order. After all, Robin Hood hashing is named for ``giving back to the poor'': it prioritizes items that are already farther from their hashes. One might expect ARCH to give a small item $x$ very large expected stretch, since a very large item $y$ with $h(y) > h(x)$ can preempt $x$ and blow up its stretch. However, the smaller $x$ is, the less likely it is that any other item $y$ hashes into the region where it would preempt $x$. As we will see in the proof of \Cref{thm:arch}, this is why ARCH achieves worst-case expected stretch $\eps^{-1}$.

\paragraph{Analysis.}
We now give a simple, self-contained proof that the ARCH table achieves worst-case expected stretch $O(\eps^{-1})$.
The result mirrors the standard results in queuing theory for LIFO-PR schedules on M/G/1 queues~\cite{Kleinrock76,WiermanHB03}. We remark, however, that the proof itself is necessarily different, because the standard approaches rely heavily on the Poisson arrivals of the M/G/1 model in order to perform an optional-stopping-time argument; to get the same result in the hash-table setting, we use an alternative approach in which we reason about $x$'s query cost by considering the impact for each item $y$ of deleting and reinserting $y$ into the table.

\thmARCH*

\begin{proof}
    Fix $x\in S$. For $T\subseteq S$ containing $x$, let $L_x(T)$ denote the scan cost of $x$ in the ARCH table on $T$, let \(e_T(x)\) be the last cell of \(x\) in that table, and let \(J_x(T) = (h(x), e_T(x)+1)\).
    We say that an item \(y \in T\) \defn{preempts} \(x\) in \(T\) if \(h(y) \in J_x(T)\).
    Then \(L_x(T)\) equals \(|x|\) plus the total length of the other items that preempt \(x\) in \(T\).

    We prove the positive-query bound by induction on $|S|$.
    If $S=\{x\}$, then $L_x(\{x\}) = |x|$.
    In general, for \(y\ne x\), consider removing \(y\) and then reinserting it.
    Then \(y\) preempts \(x\) in \(S\) exactly when \(h(y)\in J_x(S\setminus\{y\})\).
    The interval \(J_x(S\setminus\{y\})\) has length at most $L_x(S\setminus\{y\})$ and depends only on the hashes of $S\setminus\{y\}$.
    Independence of hashes therefore gives
    \[
        \Pr[h(y)\in J_x(S\setminus\{y\})]
        =\frac{\E[|J_x(S\setminus\{y\})|]}{m}
        \le\frac{\E[L_x(S\setminus\{y\})]}{m}.
    \]
    Each set \(S \setminus \{y\}\) has load at most \(1 - \eps\), so by the induction hypothesis, $\E[L_x(S\setminus\{y\})] \le \frac{|x|}{\eps}$. Therefore,
    \begin{align*}
        \E[L_x(S)]
        &= |x|+\sum_{y\in S\setminus\{x\}}|y| \cdot \Pr[h(y)\in J_x(S\setminus\{y\})] \\
        &\le |x|+\frac{1}{m}\sum_{y\in S\setminus\{x\}}|y| \cdot \E[L_x(S\setminus\{y\})] \\
        &\le |x|+(1-\eps)\frac{|x|}{\eps}
        =\frac{|x|}{\eps}.
    \end{align*}

    For negative queries, since \(\E[L_y] \le |y|/\eps\) for every \(y \in S\), \Cref{lem:stop-marks} with \(C = 1/\eps\) bounds the expected scan cost by \(1 + 1/\eps\).
\end{proof}

\begin{corollary}[Anti-Robin-Hood Catalan Hashing in External Memory]
\label{cor:intro-arch-em}
Under the working-set assumption (see \Cref{sec:preliminaries}), the hash table of \Cref{thm:arch} is cache-oblivious and answers positive queries for any item $x \in S$ in expected $1 + O(|x|/(\eps B))$ I/Os and negative queries in expected $1 + O(1/(\eps B))$ I/Os.
\end{corollary}

The ARCH hash table is already good enough for the guarantees required by the dynamic version in \Cref{sec:dynamic-high-load}.
In particular, we bound the expected query cost of any item in terms of its size, or equivalently, the \emph{maximum expected query stretch}, where the maximum is taken over all items.
\Cref{thm:arch} shows that the maximum expected query stretch in an ARCH table is at most $1/\eps$.

In the static case, we can also consider the \emph{expected maximum stretch}.
It turns out that this quantity can be quite large in ARCH tables. 
If the table contains $\Theta(n)$ items of length $2$ and one large item $y$ of size $L = \Theta(\delta n)$, then with constant probability, $c(y)$ is the cell holding the second word of some length-$2$ item $x$ (a constant fraction of all cells are such cells), so $y$ preempts $x$ after its first word and $x$ has query stretch at least $L/2 = \Omega(\delta n)$.

To bound both per-item expected query cost and expected maximum stretch across the entire table, we turn to a different fair queuing protocol, the fair-sojourn protocol, based on processor sharing.

\subsection{Fair-Sojourn Catalan Hashing (FSCH)}
\label{subsec:fsch-definition}

\defn{Processor Sharing (PS)}~\cite{Kleinrock67} is a service schedule which can be thought of as an idealized version of round-robin scheduling.
In processor sharing, if \(j\) jobs are present in the system, they all share the server equally, each receiving service at instantaneous rate \(1/j\).
Processor sharing is the canonical fair scheduling policy in queuing theory, avoiding both the starvation of large jobs (as in simple Catalan hashing) and the preemption of small jobs by newly arriving large jobs (as in ARCH)~\cite{WiermanHB03}.

Processor sharing cannot be used directly in practice, because jobs cannot truly share a server.
Similarly, we cannot write fractions of words from multiple different items in a single hash-table cell.
In practice, processor sharing is approximated in a number of different ways, the simplest of which is round-robin scheduling, which converges to PS as the size of the time quantum tends to 0~\cite{Kleinrock67}.
However, in our case, if we alternated words round-robin among active items, the resulting interleaved words would violate the Catalan nesting property (\Cref{def:catalan-property}) and could not be processed using our parser.

We will instead use the \defn{Fair-Sojourn Protocol (FSP)}~\cite{FriedmanH03}, which is a schedule that realizes the benefits of PS while running a single job on the server at each point in time.
FSP runs a virtual simulation of PS and assigns the processor to the remaining job that would finish first in this simulation.
We use the following discrete-time version, which we show approximates FSP, which in turn dominates PS.

\paragraph{The virtual PS schedule.}
Let $S$ be a set of items with arrivals $A_x = h(x)$ and service requirements $s_x = |x|$, such that $\alpha = \frac{1}{m}\sum_{x \in S} s_x = 1-\eps$.
We run a continuous processor sharing schedule: every item $x \in S$ arrives at $A_x$ and departs at a well-defined PS departure time $D_x^{\mathrm{PS}}$, satisfying
\begin{equation}
\label{eq:ps-service-integral}
    \int_{A_x}^{D_x^{\mathrm{PS}}} \frac{dt}{N_S(t)} = |x|,
\end{equation}
where the integral is interpreted cyclically and $N_S(t)$ denotes the number of jobs of $S$ alive in the virtual PS schedule at time $t$.
\Cref{eq:ps-service-integral} expresses that between \(A_x\) and \(D_x^\mathrm{PS}\), \(x\) receives service at rate \(1/N_S(t)\) at time \(t\), which is the defining property of PS.
Given $S$ and its hashes, we can compute these departure times by sorting the arrivals and simulating the schedule from an idle point.
Between consecutive arrivals and departures, $N_S(t)$ is constant, so every active item's remaining service decreases linearly at rate $1/N_S(t)$.

\paragraph{FSCH Hash Tables.}
We define the \defn{FSCH priority} $\pifsch$ by giving higher priority to items with earlier departure times in the virtual PS schedule, interpreted cyclically.
We define the \defn{FSCH table} to be $\HT_{\pifsch}$, as in \Cref{def:catalan}.

The FSCH table defines a hash-table-queue schedule which resembles a discrete-time version of FSP.
Continuous FSP uses the arrival times $h(x)$ and may preempt a job at a nonintegral time.
FSCH instead makes $x$ available at $c(x)=\lfloor h(x)\rfloor$ and assigns each occupied cell entirely to one item, so preemptions occur only between words (i.e. at integral times).
Both continuous FSP and FSCH priorities are computed from the virtual PS schedule with the original, unrounded arrivals.
As shown in \Cref{lem:fsch-rounding}, this increases sojourn time by at most one unit.

\subsection{FSCH Analysis}

We first show that the FSCH sojourn time exceeds the continuous FSP sojourn time by at most one unit.

\begin{lemma}
\label{lem:fsch-rounding}
In a hash-table queue for \(S\), let \(D_x^{\mathrm{FSP}}\) and \(D_x^{\mathrm{FSCH}}\) be the departure times of an item \(x\in S\) under continuous FSP and FSCH, respectively, with cell \(j\) representing service during \([j,j+1)\).
Then
\[
    [D_x^{\mathrm{FSCH}} - c(x)] - [D_x^{\mathrm{FSP}} - h(x)] \le 1,
\]
where times are interpreted cyclically.
\end{lemma}

\begin{proof}
Fix \(x\), write \(T = D_x^{\mathrm{FSP}}\), and let \(P_x\) consist of \(x\) and all higher-priority items arriving before \(T\).
FSP has completed every item of \(P_x\) by \(T\), since otherwise a higher-priority item would prevent \(x\) from finishing.

Rounding the arrivals of \(P_x\) down preserves this deadline.
A schedule serving \(P_x\) whenever any of its work is pending therefore also finishes by \(T\).
Its arrivals and service requirements are integers, so it finishes by \(\lfloor T\rfloor\).
In the full FSCH schedule, lower-priority items never delay \(P_x\), and every higher-priority item outside \(P_x\) has rounded arrival at least \(\lfloor T\rfloor\).
Thus \(D_x^{\mathrm{FSCH}} \le \lfloor T\rfloor \le T\), and hence
\[
    [D_x^{\mathrm{FSCH}} - c(x)] - [D_x^{\mathrm{FSP}} - h(x)]
    \le h(x) - \lfloor h(x)\rfloor < 1.
\]
\end{proof}

Next, a classical theorem of Friedman and Henderson~\cite{FriedmanH03} shows that FSP completes each item no later than PS.

\begin{theorem}[\cite{FriedmanH03}]\label{thm:fsp-ps-domination}
For any fixed arrival times and service requirements, continuous FSP and processor sharing satisfy
\[
    D_x^{\mathrm{FSP}} \le D_x^{\mathrm{PS}}
    \qquad \text{for every } x \in S,
\]
where times are interpreted cyclically.
\end{theorem}

Using \Cref{lem:fsch-rounding,thm:fsp-ps-domination}, any result on continuous PS for hash-table queues that bounds the sojourn time of an item \(x\) produces the same bound on the query scan cost of \(x\) in the FSCH table (up to rounding).
We want to bound both the expected stretch of each item and the expected maximum stretch.
Our results are analogues of similar results in the M/G/1 setting.
There, each item's expected stretch is $1/(1-\rho)$, where \(\rho\) is the load; see e.g. Kleinrock~\cite{Kleinrock76} or Harchol-Balter~\cite[Section 30.1]{HarcholBalter13}.
Cheung, van den Berg, and Boucherie~\cite{CheungVanDenBergBoucherie2006} give bounds on all moments of the conditional sojourn time that are insensitive to the job-size distribution, yielding exponential tail bounds for stretch.
These techniques do not immediately transfer to hash-table queues: an M/G/1 queue has Poisson arrivals and independently sampled service requirements, whereas hash-table queues have fixed finite sets of jobs with arbitrary sizes and independent uniform arrival times.
In particular, the numbers of arrivals in disjoint intervals are dependent, so we cannot directly apply results that assume Poisson arrivals.

Our analysis instead establishes analogous arrival and deletion properties for hash-table queues.
An arriving job sees the system with that job deleted at a uniformly random time, since its hash is independent of the other hashes.
We show that PS departure times are also independent and uniform: reversing time gives a valid PS schedule, and the map from arrivals to departures preserves measure.
Now consider the instant at which a job $x$ has received a fixed amount $v\in(0,|x|)$ of service.
Truncating $x$ to size $v$ makes this instant its departure time, allowing us to show that the number of other jobs present has the same distribution as the population of the system without $x$ at a uniformly random time.
Integrating over $v$ gives the \defn{service-deletion identity} (\Cref{lem:ps-service-deletion}).

This identity lets us bound the queue population recursively, by deleting one job at a time.
By doing so, we obtain a geometric tail bound on the number of jobs present at a random time \(u\): \(\Pr[N_S(u)\ge k]\le\alpha^k\) (\Cref{lem:ps-population-tail}).
The service-deletion identity also expresses $x$'s expected stretch as one plus the expected population of the system without $x$, giving a bound of $1/(1-\alpha)\le1/\eps$.
Finally, every item's stretch is at most the maximum queue population, since a job receives service at rate $1/j$ whenever $j$ jobs are present.
The population increases only at arrivals, so a union bound over the $n$ arrivals gives the logarithmic factor in the expected maximum stretch.
We defer the full analysis to \Cref{sec:ps-analysis}, whose main result we state here.

\begin{restatable}{theorem}{thmPSStretch}\label{thm:ps-stretch}
    Let $S$ be a nonempty set of $n$ items with $\load(S)=\alpha\le1-\eps$.
    In the PS schedule for \(S\) in the hash-table queue, each item has expected stretch at most $1/\eps$, and the expected maximum stretch is at most $(\ln n+1)/\eps$.
\end{restatable}

We now combine \Cref{thm:ps-stretch,lem:fsch-rounding,thm:fsp-ps-domination,lem:stop-marks} to establish the bounds for FSCH:

\begin{theorem}[Fair-Sojourn Catalan Hashing]
\label{thm:fsch}
Let $\eps \in (0, 1)$, and let $S$ be a  static set of $n$ items stored in an $m$-cell table with load factor $\load(S) \le 1 - \eps$. Fair-sojourn Catalan hashing (FSCH) uses $O(1)$ bits of metadata per cell, answers positive queries for any item $x \in S$ with expected scan cost at most $1 + |x|/\eps$, and answers negative queries with expected scan cost at most $2 + 1/\eps$. Moreover, if $L_x$ denotes the scan cost for $x \in S$, the expected maximum stretch across the entire table satisfies
\[ \E\left[\max_{x \in S} \frac{L_x}{|x|}\right] \le 1 + \frac{\ln n + 1}{\eps}.\]
\end{theorem}

\begin{proof}
Each cell uses the $O(1)$ bits of the Catalan representation.
By \Cref{lem:fsch-rounding,thm:fsp-ps-domination}, the positive query scan cost satisfies $L_x\le T_x+1$, where $T_x$ is $x$'s PS sojourn time.
Thus \Cref{thm:ps-stretch} gives $\E[L_x]\le 1+|x|/\eps\le(1+1/\eps)|x|$, since $|x|\ge1$.
Applying \Cref{lem:stop-marks} with $C=1+1/\eps$ gives expected scan cost at most $2+1/\eps$ for negative queries.
Since $|x|\ge1$, the expected maximum query stretch is at most
\[
    \E\left[\max_{x\in S}\frac{L_x}{|x|}\right]
    \le 1+\E\left[\max_{x\in S}\frac{T_x}{|x|}\right]
    \le 1+\frac{\ln n+1}{\eps}.
\]
\end{proof}

\section{Graveyard Catalan Hashing (Dynamic, High Load)}
\label{sec:dynamic-high-load}

In this section, we combine the two constructions developed in the preceding sections.
The hash tables in \Cref{sec:fair-catalan} achieve load factor $1-\eps$ and expected query scan cost $O(|x|/\eps)$, but are static. The dynamic table of \Cref{sec:backup-table} supports efficient updates, but its erasure codes and backup copies require a constant factor of extra space.
In the following, we will construct a fully dynamic Catalan hash table at high load factor $1 - \eps$, proving \Cref{thm:intro-dynamic-high}.

The basic idea is to keep almost all items in a static table, and reserve $\Theta(\eps m)$ interleaved cells for a dynamic table that receives new insertions. After $\Theta(\eps m)$ words of updates, we rebuild the static table on the current live set. The dynamic table's constant-factor space overhead then uses only $O(\eps m)$ cells, and a rebuild with scan cost $O(m)$ has amortized cost only $O(1/\eps)$ per word of updates.

Our construction takes inspiration from \defn{graveyard hashing}~\cite{BenderKuszmaulKuszmaul2021GraveyardHashing}, which achieves dynamic high-load operations in unit-size linear probing by maintaining tombstones from deletions (and inserting artificial tombstones during periodic rebuilds) so that new insertions quickly find tombstones without scanning to the end of a long cluster. 
In our variable-length external-memory setting, graveyard hashing would be subject to the same write amplification issue we described in \Cref{sec:erasure-oracle}. Moreover, both the ARCH and FSCH schemes in \Cref{sec:fair-catalan} use an inherently static scheduling-based approach.

Instead, we implement the graveyard hashing framework in a more heavy-handed structural manner: we explicitly decompose the hash table into interleaved tables that share the same physical array.
Into a large static table, we sparsely interleave a small dynamic table that serves the same function as the tombstones in graveyard hashing: it efficiently accepts new insertions.
Also as in graveyard hashing, periodically we rebuild to refresh the dynamic table.

In the next sections, we describe the construction and its operation bounds; we later show how to rebuild in a constant number of sequential passes (\Cref{sec:graveyard-rebuild}).

\subsection{A Static Table with a Small Dynamic Buffer}
\label{sec:graveyard-architecture}

We divide the operation sequence into \defn{rebuild windows}. At the start of a window, all live items are stored in a static table $\HT_{\mathrm{full}}$ with $m$ cells. During the window, insertions go into a small dynamic table $\CHT_{\mathrm{small}}$, and deletions from $\HT_{\mathrm{full}}$ merely set a flag. Thus the static layout remains fixed until the next rebuild.

\paragraph{The Three Tables.}
Let $\eps'=\eps/C$, where $C$ is a sufficiently large constant, and set
\[
    P=\lfloor 1/\eps'\rfloor.
\]
Let $m_{\mathrm{small}}$ be the smallest power of two at least $\lceil m/P\rceil$, so that the auxiliary tables satisfy the size convention of \Cref{sec:backup-construction}.
The construction consists of the following tables.
\begin{enumerate}[label=(\arabic*)]
    \item \textbf{The static table $\HT_{\mathrm{full}}$.} This is an $m$-cell ARCH table built on the live set $T$ at the start of the window.
    We keep $T$ fixed throughout the window, including items subsequently marked as deleted.
    (We fix the use of ARCH for ease of exposition. Since FSCH guarantees the same expected query scan bounds, all of the claimed bounds in this section would hold with a FSCH-based static table as well.)
    We add two additional metadata bits to each cell: a \defn{deletion bit}, which marks the first cell of an item to indicate it's deleted, and a \defn{collision bit}, defined below, which enables queries for deleted items to resolve fingerprint collisions.
    Deleted items' words remain in place.
    The static table's placement rule and remaining metadata are unchanged.
    \item \textbf{The insertion table $\CHT_{\mathrm{small}}$.} This is the dynamic low-load structure
    \[
        \CHT_{\mathrm{small}}
        =\CHT(\HT_{\mathrm{small}},\member_{\mathrm{small}},\backup_{\mathrm{small}})
    \]
    of \Cref{sec:backup-table}, extended to support deletions by \Cref{lem:low-load-deletions} below. Each of its three constituent tables has $m_{\mathrm{small}}$ cells. It starts empty and stores the items inserted during the window that are still live.
    As in \Cref{sec:backup-table}, we sometimes abuse notation and refer to a triple as a cell, for example, the canonical cell \(c_{\mathrm{small}}(x)\) of an item \(x\).
    \item \textbf{The deletion table $\DHT$.} This is an ordinary insertion-only unordered linear-probing table with $m_{\mathrm{small}}$ cells. It starts empty. Whenever an item of $\HT_{\mathrm{full}}$ is deleted, we insert a 1-word descriptor $d(k)$ into \(\DHT\), starting at $c_{\mathrm{small}}(k)$.
    \(d(k)=(c(k),\psi(k))\) consists of \(k\)'s canonical cell together with a \(4\log{m}\)-bit fingerprint from an independent hash function.
    During rebuilding, we set the collision bit at a static cell $j$ exactly when two distinct keys in $T$ with canonical cell $j$ have the same fingerprint.
    These bits remain fixed throughout the window.
    Thus, when the collision bit at $c(k)$ is clear, at most one key in $T$ has descriptor $d(k)$.
    We always insert one record per deleted static item, even in the above case where two descriptors are the same, allowing equal descriptors to occupy separate cells.
    Records stay in $\DHT$ until the next rebuild, even if their keys are subsequently reinserted into $\CHT_{\mathrm{small}}$.
\end{enumerate}
At all times, each live item occurs either as an unflagged item of $\HT_{\mathrm{full}}$ or as a live item of $\CHT_{\mathrm{small}}$, and $\DHT$ contains one descriptor for each flagged item of $\HT_{\mathrm{full}}$.

The deletion table is structured using descriptors with fingerprints to handle a complication that can arise with deletions.
A negative query for a deleted key could still scan the entire old item, which may be much longer than the key itself.
The static negative-query bound does not apply, because this key's hash is correlated with the contents of the static table.
Storing the deleted key in a second table would reduce this cost only to one proportional to the key's length, but we reduce this further to \(O(1/\eps)\) in expectation by using constant-size descriptors.
We check for collisions among resident keys during rebuilding, so that a fingerprint match can certify absence without reading the deleted key.

\paragraph{Interleaving and Space.}
As in \Cref{sec:backup-reduction}, we interleave all constituent tables in a single physical array.
When interleaving the cell $s$ (technically three cells) of $\CHT_{\mathrm{small}}$, we follow it with cell $s$ from \(\DHT\).
We refer to this collection of cells as a single \defn{small cell} $s$.
For $0\le s<m_{\mathrm{small}}$, place small cell $s$ immediately after cell $\lceil (s+1)m/m_{\mathrm{small}}\rceil-1$ of $\HT_{\mathrm{full}}$.
Thus consecutive small cells are separated by at most $\lceil m/m_{\mathrm{small}}\rceil\le P$ static cells and four component cells of the small cell.

Assume for now that $\eps'm\ge 17$. Since $\eps'<1/2$, we have $m_{\mathrm{small}}\ge\eps'm$ and $m_{\mathrm{small}}<2\lceil m/P\rceil\le 2(2\eps'm+1)$, so
\[
    m_{\mathrm{total}}=m+4m_{\mathrm{small}} \le (1+28\eps')m.
\]
Choosing $C\ge 56$ ensures that a live volume of at most $(1-\eps)m_{\mathrm{total}}$ is at most $(1-\eps/2)m$.
In particular, each rebuilt static table has slack at least $\eps/2$.
Consecutive cells of any component of the small table are at most
\(
    \sigma=P+4=O(1/\eps)
\)
physical cells apart.

All tables use the same hash, rescaled to their lengths. A key $k$ has canonical cell $c(k)=\lfloor h(k)\rfloor$ in the static table and
\[
    c_{\mathrm{small}}(k)
    =\left\lfloor \frac{m_{\mathrm{small}}}{m}h(k)\right\rfloor
\]
in the small table.
\(\csmall(k)\) lies within $O(P)$ physical cells after $c(k)$.
Consequently, scans in all three structures start near the same physical location, which preserves the I/O bounds.

\paragraph{The Update Budget.}
We keep two counters: the total volume inserted into $\CHT_{\mathrm{small}}$, including items subsequently deleted, and the total length of keys deleted from $\HT_{\mathrm{full}}$. The latter also bounds the number of records inserted into $\DHT$. We ensure that neither counter ever exceeds
\(
    b=\eps'm/64\le m_{\mathrm{small}}/64.
\)
Before an operation would exceed either budget, we rebuild and then perform the operation in the new window. An insertion of length at least $b/2$, or a deletion from $\HT_{\mathrm{full}}$ whose key has length at least $b/2$, is instead incorporated directly into a rebuild. We call these \defn{long updates}. This rule keeps the insertion table at the low load required by \Cref{sec:backup-table}, and keeps $\DHT$ at load at most $1/64$. It also ensures that a budget-triggered rebuild follows at least $b/2=\Omega(\eps m)$ words charged to one of the counters.

If $\eps'm<17$, we use just the static table and rebuild after every update. Then $m=O(1/\eps)$, so a rebuild has scan cost $O(1/\eps)$; under the cache-size assumption in \Cref{sec:graveyard-bounds}, the whole table fits in cache. The bounds below therefore also cover this case.

\subsection{Operations Within a Rebuild Window}
\label{sec:graveyard-operations}

We use the low-load dynamic table from \Cref{thm:intro-dynamic-low} as a black box, except we extend it to support deletions.

\begin{lemma}[Low-Load Deletions]
\label{lem:low-load-deletions}
The structure $\CHT(\HT,\member,\backup)$ of \Cref{sec:backup-table}, with $m$ cells in each constituent table and initially empty, supports any operation sequence whose insertions have total volume at most $m/64$. Its scan costs are:
\begin{enumerate}[label=(\arabic*)]
    \item $O(|x|)$ in amortized expectation for a positive query for an item $x$;
    \item $O(1)$ in expectation for a negative query; and
    \item $O(|x|\log^{1+\delta}|x|)$ in amortized expectation for an insertion or deletion of an item $x$.
\end{enumerate}
Each operation on a key $k$ makes $O(1)$ forward passes over a contiguous physical interval starting at $c(k)$. Under the cache assumptions of \Cref{sec:backup-table}, the corresponding I/O bounds are $1+O(L/B)$, where $L$ is the stated scan bound.
\end{lemma}

The proof is given in \Cref{app:deletions}. The main issue is repeated deletion and reinsertion of one key: its successive values share a hash, so they cannot be treated as independent items. The construction reuses a deleted copy when that key is reinserted into the same size class. We use the budget $m/64$ so that the density bound for the backup table continues to apply.

\begin{lemma}[Interleaving Preserves Scan Bounds]
\label{lem:strided-embedding}
An operation in $\CHT_{\mathrm{small}}$ or $\DHT$ with logical scan bound $L$ has physical scan bound $O(\sigma(L+1))$. Its accesses make $O(1)$ forward passes over an interval beginning within $O(\sigma)$ cells after $c(k)$ in $\HT_{\mathrm{full}}$. The same statement holds for expected and amortized expected bounds.
\end{lemma}
\begin{proof}
An interval of $L$ cells in any auxiliary constituent table spans at most $\sigma(L+1)$ physical cells. All constituent scans start at their canonical cells. For $g=c_{\mathrm{small}}(k)$, the boundary $(g+1)m/m_{\mathrm{small}}$ lies in $(h(k),h(k)+m/m_{\mathrm{small}}]$. Rounding this boundary changes its position by less than one static cell, and $m/m_{\mathrm{small}}\le P$. Thus these starting points lie within $O(P)$ physical cells after $c(k)$. The encoding, decoding, and allocator bookkeeping use the same scratch region as before, so interleaving changes only the lengths of the scans.
\end{proof}

\paragraph{Deletion-Table Scans.}
A search in $\DHT$ starts at $c_{\mathrm{small}}(k)$ and scans to a matching descriptor or the first empty cell. An insertion scans to the first empty cell and writes its record there. Each static item is deleted at most once per window. Moreover, the window boundaries and the sequence of deleted static keys are determined by the operation sequence, independently of the hashes. Thus the records have independent uniform home small cells, even if their descriptors coincide. Since the load is at most $1/64$, ordinary unit-size linear probing gives expected logical scan cost $O(1)$ for either operation, whether or not the query key has a record. By \Cref{lem:strided-embedding}, the physical scan cost is $O(1/\eps)$ in expectation.

\paragraph{Queries.}
A query for $k$ reads the collision bit at $c(k)$ and scans forward from that cell, running the following searches together:
\begin{itemize}
    \item Search $\CHT_{\mathrm{small}}$ for a live item with key $k$.
    \item If the collision bit is clear, search $\DHT$ for $d(k)$.
    \item Search $\HT_{\mathrm{full}}$ with the Catalan parser, treating flagged items as absent. Abandon this search as soon as the deletion-table search finds $d(k)$.
\end{itemize}
Return the live item if one is found, and otherwise return \textsc{Not Found} once the insertion-table search and either the static-table search or the deletion-table search establish absence. The searches advance together through the interleaved array; any backup lookup or reinflation uses the forward passes of the low-load construction.

To see why a deletion-table match certifies absence from the live static set, let $z\in T$ be the deleted item that supplied the matching descriptor. If $k$ were also the key of a live static item $x$, then $d(k(x))=d(k(z))$. Since the collision bit at $c(k)$ is clear, $x=z$, a contradiction. A collision with a key outside $T$ is harmless: that key is absent from the static table anyway. The query must still search the insertion table, since a deleted static key may have been reinserted there.

\begin{lemma}[Query Bounds Within a Window]
\label{lem:graveyard-queries}
A positive query for an item $x$ has amortized expected scan cost $O(|x|/\eps)$. Every negative query has expected scan cost $O(1/\eps)$, independently of the key length. Each query makes $O(1)$ forward passes over a contiguous physical interval starting at $c(k)$.
\end{lemma}
\begin{proof}
If the live item $x$ is in the insertion table, its search finishes with amortized expected scan cost $O(|x|/\eps)$ by \Cref{lem:low-load-deletions,lem:strided-embedding}; the concurrent searches stop no later. If $x$ is an unflagged static item, the static search has expected scan cost $O(|x|/\eps)$, while the insertion-table search is negative and costs $O(1/\eps)$ in expectation.

Now consider a negative query for $k$. The insertion-table search has expected scan cost $O(1/\eps)$. If $k$ is not a key of $T$, its hash is independent of the static contents, so the static search has expected scan cost $O(1/\eps)$ by \Cref{thm:arch}; an early stop from $\DHT$ can only shorten it.

Otherwise, $k$ belongs to a flagged static item. Let $\mathcal E$ be the event that two distinct keys of $T$ have equal fingerprints. Since $|T|\le m$,
\[
    \Pr[\mathcal E]\le \binom{|T|}{2}m^{-4}=O(m^{-2}).
\]
Outside $\mathcal E$, the collision bit at $c(k)$ is clear and the query stops the static search upon finding $d(k)$ in $\DHT$. Let $L_{\DHT}$ be the physical distance from $c(k)$ through this record, with $\E[L_{\DHT}]=O(1/\eps)$ by the deletion-table scan bound. The static and deletion-table searches together cost at most
\[
    O(L_{\DHT})+O(m)\ind[\mathcal E].
\]
Taking expectations gives $O(1/\eps)+O(m)\Pr[\mathcal E]=O(1/\eps)$, without conditioning the deletion-table bound on the collision bits. This estimate applies separately to every query, including repeated queries for the same deleted key; no query cost is charged to its deletion.
\end{proof}

\paragraph{Insertions.}
To insert an item $x$ whose key is absent, first apply the budget rule. A long insertion is incorporated directly into a rebuild; otherwise, rebuild first if necessary, and then insert $x$ into $\CHT_{\mathrm{small}}$. Its key may already have a dead copy there; \Cref{lem:low-load-deletions} handles this case. Apart from rebuilding, \Cref{lem:low-load-deletions,lem:strided-embedding} give amortized expected scan cost $O(|x|\eps^{-1}\log^{1+\delta}|x|)$.

\paragraph{Deletions.}
To delete a present key $k$, first locate its item $x$ with a query. If $x$ is in $\CHT_{\mathrm{small}}$, delete it using \Cref{lem:low-load-deletions}. Otherwise, check the budget rule before changing the table. A long deletion is incorporated directly into a rebuild. If the deletion is short but the key budget would be exceeded, rebuild first and locate $x$ in the rebuilt static table. Then set the deleted flag at its opening cell and insert $d(k)$ into $\DHT$. The static table's delimiters, home bits, stop marks, and collision bits remain unchanged, so its parser continues to operate on exactly the layout built at the start of the window. The auxiliary update has expected scan cost $O(1/\eps)$, which is within the claimed update bound.

Thus both insertions and deletions have the desired update bound within a window. It remains to pay for rebuilding; we defer the analysis of the amortized expected cost of rebuilds to \Cref{sec:graveyard-rebuild}.

\subsection{Putting the Bounds Together}
\label{sec:graveyard-bounds}

\paragraph{Cache Size for I/O Bounds.}
Our I/O bounds use the working-set assumption of \Cref{sec:preliminaries}, together with $M \ge C_M(B + s_{\max}\eps^{-2}\log^3 m)$ for a sufficiently large constant $C_M$. This accommodates the encoding, decoding, and bookkeeping space of \Cref{sec:backup-table}, as well as $O(s_{\max}\eps^{-2}\log^3 m)$ words of workspace for rebuilding one run at a time in cache (\Cref{sec:graveyard-rebuild}). We remark that our assumption on $M$ is quite mild in the sense that it only requires memory to be slightly larger than what is required to fit the largest string.

The following proposition collects the guarantees established by the construction, proving \Cref{thm:intro-dynamic-high,cor:intro-dynamic-high-em}.

\begin{proposition}[Dynamic High-Load Bounds]
\label{prop:graveyard-bounds}
For $\eps\in(0,1/2)$ and fixed $\delta>0$, the construction operates at load factor $1-\eps$ and has the following amortized expected scan costs:
\begin{enumerate}[label=(\arabic*)]
    \item $O(|x|/\eps)$ for a positive query for an item $x$;
    \item $O(1/\eps)$ for a negative query, independently of the key length; and
    \item $O(|x|\log^{1+\delta}|x|/\eps)$ for an insertion or deletion of an item $x$.
\end{enumerate}
Under the cache-size assumption above, the construction is cache-oblivious and the corresponding amortized expected I/O bounds are
\[
\begin{array}{ll}
    \text{positive query:} & 1+O\!\left(|x|/(\eps B)\right),\\[2pt]
    \text{negative query:} & 1+O\!\left(1/(\eps B)\right),\\[2pt]
    \text{insertion or deletion:} & 1+O\!\left(|x|\log^{1+\delta}|x|/(\eps B)\right).
\end{array}
\]
\end{proposition}
\begin{proof}
The space calculation in \Cref{sec:graveyard-architecture} leaves slack at least $\eps/2$ in the static table. The update budgets keep the insertion table within the hypotheses of \Cref{lem:low-load-deletions} and keep the deletion table at load at most $1/64$. Thus \Cref{lem:graveyard-queries} and the update analysis give all within-window scan bounds.

Each budget-triggered rebuild follows $\Omega(\eps m)$ words of insertions or deleted keys. Charge its $O(m)$ expected scan cost to those words. A long update has length $\Omega(\eps m)$ and pays for its own rebuild. Since a deleted key is no longer than its item, rebuilding adds only $O(|x|/\eps)$ to each update's amortized expected scan cost.

For I/Os, all within-window accesses make $O(1)$ forward passes over a contiguous interval starting at $c(k)$. The cache-size assumption above converts a scan bound $L$ to $1+O(L/B)$ expected I/Os, with a single leading $1$ shared by the interleaved searches. By \Cref{lem:streaming-rebuild}, rebuilding adds $O(|x|/(\eps B))$ amortized expected I/Os per update. None of the algorithms uses $M$ or $B$.
\end{proof}

\section*{Acknowledgments}
This work is partially supported by NSF grants
2504470, 2504471, 2551615, and 2542165, as well as by a CMU / Jane Street collaboration grant.

\section*{AI Disclosure}
All of the main technical ideas in this paper were generated by the authors over the duration of around a year; during that time, the authors used AI tools (all models of GPT, Claude, and Gemini during that time span) to aid in literature review and to explore different ways of structuring the proofs. AI tools were also used in the writing process to help write the starting draft for the paper, which was then almost entirely rewritten by the authors. Finally, AI tools were used to polish text throughout, and to perform edits to the paper.

We reviewed and revised the AI-assisted material and take full responsibility for the content, correctness, and originality of the paper.

\printbibliography

\appendix
\crefalias{section}{appendix}
\section{The Static Catalan Parser}\label{app:parser}

\paragraph{State.}
The parser maintains a stack $\mathcal{S}$ of active item frames for items whose \copen{} tag has been seen at or after cell $t$, along with a single integer counter $f \ge 0$ (initially $0$) that counts observed words belonging to an ``unmatched suffix'' of an item whose \copen{} tag occurred \emph{before} cell $t$. 
Each frame on $\mathcal{S}$ buffers the key, maintaining a word counter $cnt$ (the number of words of that item seen so far) and a boolean match flag indicating whether the key words seen so far match the corresponding prefix of the queried key $k$.
When the buffer reaches the end of the key, it computes and retains its canonical cell before discarding the buffer.
The match flag tests equality of the complete keys, including their lengths.
If the key matches $k$, the frame buffers the subsequent value in order to return it.
These key and value buffers are part of the parser state assumed to reside in cache in \Cref{sec:preliminaries}.

\paragraph{Transitions on Reading a Cell $j$.}
\begin{enumerate}[label=(\roman*)]
    \item \copen{}: Push a new frame onto $\mathcal{S}$ with $cnt = 1$, and compare the cell's word with the first word of $k$.
    \item \copenclose{}: This cell is a complete $1$-word item; if its key is $k$, return \textsc{Found} (we assume a 1-word item has no value).
    Otherwise, compute its canonical cell and apply the stop-cell rule of \Cref{sec:catalan-query}.
    The stack $\mathcal{S}$ and counter $f$ are unchanged.
    \item \cmiddle{} with $\mathcal{S}$ non-empty: Increment $cnt$ of the top frame of $\mathcal{S}$, and, if the $cnt$-th word of the item is part of its key, compare it with the $cnt$-th word of $k$.
    \item \cclose{} with $\mathcal{S}$ non-empty: Increment $cnt$ of the top frame of $\mathcal{S}$, compare the cell's word with $k$ as in (iii), and pop the frame. If the completed item's key is $k$, return \textsc{Found}, together with its buffered value. Otherwise, use its retained canonical cell and the stop bit of cell $j$ to apply the stop-cell rule of \Cref{sec:catalan-query}.
    \item \cmiddle{} with $\mathcal{S}$ empty: Increment the unmatched-suffix counter $f \gets f + 1$.
    \item \cclose{} with $\mathcal{S}$ empty: Increment $f \gets f + 1$ (completing the unmatched suffix of some item that opened before $t$), and reset $f \gets 0$.
    No stop-cell test is needed for this item: when $t=c(k)$, an item that opened before $t$ cannot have canonical cell $c(k)$.
    \item \cempty{}: Terminate the scan and return \textsc{Not Found}.
\end{enumerate}

\begin{lemma}[Correctness of the Catalan Parser]
\label{lem:parser-correct}
Suppose a table satisfies the Catalan nesting property (\Cref{def:catalan-property}), and a scan starts at cell $t$. Then the Catalan parser correctly assembles, in logical word order, every item $y$ whose entire interval $I_y$ lies at or after $t$. Furthermore, at every step of the scan:
\begin{enumerate}[label=(\arabic*)]
    \item Each frame on the stack $\mathcal{S}$ tracks a single item $y$ that opened at or after $t$, and its counter $cnt$ equals the number of words of $y$ seen so far (so $cnt \le |y|$).
    \item Whenever $f > 0$, the unmatched-suffix counter $f$ counts words belonging to a single item $w$ that opened before $t$ (namely, the innermost such item that has not yet closed), and never combines words from two different items (so $f \le |w|$).
\end{enumerate}
\end{lemma}

\begin{proof}
    By \Cref{def:catalan-property}, whenever an item $z$ is nested inside another item $y$ ($I_z \subsetneq I_y$), $I_z$ lies strictly between two consecutive cells of $y$. If $y$ opened at or after $t$, then $y$'s frame is on the stack $\mathcal{S}$ when $z$'s \copen{} is encountered; $z$'s frame is pushed above $y$, absorbs all words of $z$ (and any items nested inside $z$), and is popped on $z$'s \cclose{} before the next word of $y$ is encountered. Thus $y$'s frame receives precisely the words of $y$ in order. Similarly, when $\mathcal{S}$ is empty, any \cmiddle{} or \cclose{} cell belongs to the innermost enclosing item $w$ that opened before $t$. Any item $z$ nested between two words of $w$ after $t$ pushes a frame onto $\mathcal{S}$ and pops it before the next word of $w$, leaving $f$ untouched during $I_z$; when $w$'s \cclose{} is reached with $\mathcal{S}$ empty, $f$ is reset to $0$ before any outer enclosing item's words can be seen. Thus $f$ never conflates words from two different items.
\end{proof}

\section{Analysis of Processor Sharing for Hash-Table Queues}
\label{sec:ps-analysis}

In this appendix, we prove \Cref{thm:ps-stretch}, which bounds the expected stretch of each item and the expected maximum stretch in the virtual processor-sharing (PS) schedule used by FSCH.
Recall the hash-table-queue model of \Cref{subsec:scheduling-view}: the set $S$ of $n$ items and their sizes are fixed, while the arrival times $A_x=h(x)$ are independent and uniform on a circle of circumference $m$.
The load is $\alpha=\frac{1}{m}\sum_{x\in S}|x|\le1-\eps$.
We write $D_x$ for the PS departure time of $x$ and $N(t)=N_S(t)$ for the number of jobs present at time $t$.
Each active job receives service at rate $1/N(t)$, and its stretch is its sojourn time divided by its size.
All times and intervals below are interpreted cyclically.

Our main tool is the service-deletion identity (\Cref{lem:ps-service-deletion}), which relates the population seen by a job during its service to the population of the system with that job removed.
We first establish the arrival and departure properties needed to prove this identity, using time reversal and a measure-preserving map from arrivals to departures.
The identity then yields a geometric tail bound on the queue population and the expected stretch bound for each item.
Finally, a union bound over arrivals controls the maximum population, giving the bound on the expected maximum stretch.

We begin by considering what an arriving job sees.
In the M/G/1 setting, PASTA (``Poisson Arrivals See Time Averages'') equates the distribution seen by arrivals with the stationary time-average distribution~\cite[Section 13.3]{HarcholBalter13}.
Here $A_x$ is uniform and independent of the hashes of $S\setminus\{x\}$, so the corresponding comparison is with the system in which $x$ is absent.

\begin{observation}\label{obs:ps-arrival}
    In a hash-table queue, the state of the system immediately before an item $x$ arrives has the same distribution as the state of the system without $x$ at a uniformly random time.
\end{observation}

We would like a similar deletion property at departures, so that it can later be applied at intermediate stages of a job's service.
Our first ingredient is time reversal, which is classical in queuing theory~\cite{Kelly79}.
We observe that this holds immediately for processor sharing in hash-table queues.

\begin{observation}\label{obs:ps-reversibility}
    In a hash-table queue, given a PS schedule with arrivals \((A_x)_{x\in S}\) and departures \((D_x)_{x\in S}\), the reverse schedule defined by \(A_x^{'} = m-D_x\) and \(D_x^{'} = m-A_x\) is a valid PS schedule.
\end{observation}

To invoke \Cref{obs:ps-arrival} in the reversed schedule, we also need its arrivals (departures in the forward schedule) to be independent and uniform.
This is the analogue of the Poisson-output property of stationary M/G/1-PS queues: in both models, departures retain the distributional form of arrivals~\cite[Chapter 22]{HarcholBalter13}.

\begin{lemma}\label{lem:ps-departures-uid}
    In a hash-table queue, the departures in a PS schedule are uniformly and independently distributed.
\end{lemma}

\begin{proof}
    Each arrival \(A_x\) and departure \(D_x\) lives in \(\R/m\Z\).
    Therefore the collection of all arrivals (resp. all departures) lives on \((\R/m\Z)^{|S|}\).
    The proof will analyze the map \(\Phi\colon (A_x)_{x \in S} \mapsto (D_x)_{x\in S}\).
    We will show that \(\Phi\) is bijective and measure-preserving.

    By \Cref{obs:ps-reversibility}, the reverse schedule is a valid PS schedule.
    Thus departures determine arrivals uniquely by reversing time and running PS again, so $\Phi$ is a bijection.

    Now, outside the measure-zero set where two or more arrivals or departures coincide, the number of jobs \(N(t)\) is locally constant except at the points \(\{A_x, D_x\}_{x\in S}\).
    We differentiate the service equations (\Cref{eq:ps-service-integral})
    \begin{equation}\label{eq:ps-service-integral-uid}
        |x| = \int_{A_x}^{D_x} \frac{dt}{N(t)}, \qquad x \in S,
    \end{equation}
    yielding \(\Gamma^A dA + \Gamma^D dD = 0\), where \(\Gamma^A\) and \(\Gamma^D\) are locally constant matrices (except at \(\{A_x, D_y\}_{x\in S}\)).
    Starting at an idle point, we order the jobs by arrival time.
    Then, for $y\ne x$, an infinitesimal perturbation of \(A_y\) only affects the integral in \Cref{eq:ps-service-integral-uid} if \(A_y\) is in the interval \((A_x, D_x)\) (interpreted cyclically).
    Therefore \(\Gamma^A\) is upper triangular with diagonal entries \(\Gamma_{xx}^A = -\frac{1}{N(A_x+)}\).
    Similarly, ordering the rows and columns by departure time makes $\Gamma^D$ lower triangular with diagonal entries $\Gamma_{xx}^D = 1/N(D_x-)$.
    This reordering does not change its determinant.
    Therefore, the Jacobian determinant of $\Phi$ is
    \[
        \left|\det \frac{\partial D}{\partial A}\right| = \frac{|\det \Gamma^A|}{|\det \Gamma^D|} = \frac{\prod_{x \in S} N(D_x-)}{\prod_{x \in S} N(A_x+)}.
    \]
    
    Every arrival where \(N(A_x+) = r\) is an upcrossing from \(r-1\) to \(r\), and every departure where \(N(D_x-) = r\) is a downcrossing from \(r\) to \(r-1\). 
    The number of upcrossings must equal the number of downcrossings, so the factors of the numerator and denominator are the same.
    Hence, $|\det(\partial D / \partial A)| = 1$ almost everywhere, and so \(\Phi\) is measure-preserving.
    Because the arrivals are u.i.d. and this is preserved by \(\Phi\), the departures are as well.
\end{proof}

Combining time reversal with \Cref{lem:ps-departures-uid} now turns the arrival observation into a departure observation.

\begin{corollary}\label{cor:ps-departure}
    In a PS schedule for a hash-table queue, the number of other jobs present immediately before a job $x$ departs has the same distribution as the number of jobs in the system without $x$ at a uniformly random time.
\end{corollary}

\begin{proof}
    This follows immediately from \Cref{lem:ps-departures-uid,obs:ps-arrival,obs:ps-reversibility}.
\end{proof}

To bound sojourn times, we need to relate the population throughout a job's service to the population of the system without that job.
For PS, truncating a job $x$ to size $v$ leaves its service history unchanged until it has received $v$ units of service, turning that instant into a departure.
Applying \Cref{cor:ps-departure} to this truncated job and integrating over $v$ gives the following identity.

\begin{lemma}[Service-Deletion Identity]\label{lem:ps-service-deletion}
    Let \(x\) be a tagged job in a hash-table queue with a PS schedule.
    Let \(\chi_x(t)\) be the indicator that \(x\) is in service at time \(t\) and \(M_x(t) = M_x^S(t) = N(t) - \chi_x(t)\) be the number of jobs present other than \(x\).
    For every function $g\colon\N\to\R$ we have
    \begin{equation}
    \label{eq:ps-deletion}
        \E\left[\int_0^m \chi_x(t) \frac{g(M_x(t))}{N(t)} \, dt\right] = \frac{|x|}{m} \E\left[\int_0^m g\big(N_{S \setminus \{x\}}(t)\big) \, dt\right].
    \end{equation}
\end{lemma}

We use \Cref{lem:ps-service-deletion} in two ways.
Taking $g(i) = i+1$ makes the left-hand side of \Cref{eq:ps-deletion} equal to the expected sojourn time of $x$.
Taking $g(i) = \ind\{i \ge k\}$ makes it the expected amount of service that $x$ receives while at least $k$ other jobs are present.
Summing this second identity over $x$ will give a tail bound on the number of jobs in the system.

\begin{proof}
    We will change variables in the integral so that we integrate against the service time attained by \(x\).
    Define the service time
    \[
        C_x(t) = \int_{A_x}^t \frac{du}{N(u)}.
    \]
    For \(v \in (0, |x|)\), define \(T_x(v)\) to be the unique time at which \(C_x(T_x(v)) = v\).
    We make the change of variables \(v = C_x(t)\), for which \(dv = \frac{dt}{N(t)}\).
    Then we have
    \[
        \int_0^m \chi_x(t) \frac{g(M_x(t))}{N(t)} \, dt = \int_0^{|x|} g\big(M_x(T_x(v))\big) \, dv.
    \]

    Now fix any \(v \in (0, |x|)\) and consider a new system \(S_v\) where \(x\) is replaced with the truncated job \(x_v\) of size \(v\).
    The systems are identical up until $T_x(v)$, when $x_v$ departs from $S_v$.
    Therefore,
    \[
        M_x^S(T_x(v)-) = M_{x_v}^{S_v}(D_{x_v}-).
    \]
    Now, applying \Cref{cor:ps-departure} to \(x_v\) in \(S_v\) shows that \(M_{x_v}^{S_v}(D_{x_v}-)\) has the same distribution as \(N_{S\setminus\{x\}}(u)\) where \(u\) is a uniformly random time.
    Taking expectations and integrating gives
\[
    \E\left[\int_0^m \chi_x(t) \frac{g(M_x(t))}{N(t)} \, dt\right] = \int_0^{|x|} \E\left[g(N_{S \setminus \{x\}}(u))\right] \, dv = \frac{|x|}{m} \E\left[\int_0^m g\left(N_{S \setminus \{x\}}(t)\right) \, dt\right].
\] 
\end{proof}

We now can prove bounds on probability there are at least a given number of jobs present at a random time with a recursion derived from summing the service-deletion identity over jobs.
The resulting upper bound mirrors the geometric stationary population law of M/G/1-PS queues, whose tail at load $\rho$ is $\rho^k$, independently of the service-size distribution~\cite[Chapter 22]{HarcholBalter13}.

\begin{lemma}
\label{lem:ps-population-tail}
    Consider a PS schedule for hash-table queue for a set \(S\) with \(\load(S) \le \alpha\).
    For a uniformly random time $u$, let $H_S(k) = \Pr[N_S(u) \ge k]$.
    Then, for every integer $k \ge 0$,
    \[
        H_S(k) \le \alpha^k \quad \textrm{and} \quad \E[N_S(u)] \le \frac{\alpha}{1-\alpha}.
    \]
\end{lemma}

\begin{proof}
    For each subset $X \subseteq S$, write $\alpha_X = \frac{1}{m}\sum_{x\in X}|x|$.
    We prove $H_X(k) \le \alpha_X^k$ simultaneously for all $X \subseteq S$ by induction on $k$.
    The base case $k=0$ holds since $H_X(0)=1$.
    Whenever $N_X(t) \ge k+1$, each active job contributes $1/N_X(t)$ to the indicator of this event. Thus
    \[
        \ind\{N_X(t) \ge k+1\}
        = \sum_{x\in X} \chi_x^X(t)\frac{\ind\{M_x^X(t) \ge k\}}{N_X(t)},
    \]
    where a summand is taken to be zero when $x$ is inactive.
    Integrating, applying \Cref{lem:ps-service-deletion} with $g(i)=\ind\{i\ge k\}$, and using the induction hypothesis gives
    \[
        H_X(k+1) = \sum_{x\in X}\frac{|x|}{m} H_{X\setminus\{x\}}(k) \le \sum_{x\in X}\frac{|x|}{m}\alpha_{X\setminus\{x\}}^k \le \alpha_X^{k+1}.
    \]
    This completes the induction. Finally,
    \[
        \E[N_S(u)] = \sum_{k\ge1} H_S(k) \le \sum_{k\ge1}\alpha^k = \frac{\alpha}{1-\alpha}.
    \]
\end{proof}

Next, we translate the bounds of the number of jobs present into stretch bounds.
The service-deletion identity makes the expected stretch of a tagged item equal to one plus the expected population without that item, yielding the same $1/(1-\rho)$ dependence on load as in M/G/1-PS~\cite[Section 30.1]{HarcholBalter13}.
For the maximum stretch, every job's service rate is at least the reciprocal of the maximum population.
The population can increase only at arrivals, so a union bound over the $n$ arrivals supplies the additional logarithmic factor.

\thmPSStretch*

\begin{proof}
    Write $T_x=D_x^{\mathrm{PS}}-A_x$ for the sojourn time of $x$.
    Applying \Cref{lem:ps-service-deletion} with $g(i)=i+1$ and then \Cref{lem:ps-population-tail}, we obtain
    \[
        \E[T_x]
        = |x|\cdot\E[N_{S\setminus\{x\}}(u)+1]
        \le |x|\left(\frac{\alpha}{1-\alpha}+1\right)
        \le \frac{|x|}{\eps}.
    \]

    Let $K=\max_t N_S(t)$. Since every active item receives service at rate at least $1/K$, we have $T_x/|x|\le K$ for every $x$.
    The population increases only at arrivals. By \Cref{obs:ps-arrival,lem:ps-population-tail}, for every integer $k\ge0$,
    \[
        \Pr[K\ge k+1]
        \le \sum_{x\in S}\Pr[N_{S\setminus\{x\}}(u)\ge k]
        \le n\alpha^k \le ne^{-\eps k}.
    \]
    Since $u\mapsto\min\{n,1/u\}$ is nonincreasing, integration over the intervals $[\alpha^{k+1},\alpha^k]$ gives
    \[
        (1-\alpha)\sum_{k\ge0}\min\{1,n\alpha^k\}
        \le \int_0^1\min\{n,1/u\}\,du = \ln n+1.
    \]
    Consequently,
    \[
        \E\left[\max_{x\in S}\frac{T_x}{|x|}\right]
        \le \E[K]
        \le \sum_{k\ge0}\min\{1,n\alpha^k\}
        \le \frac{\ln n+1}{1-\alpha}
        \le \frac{\ln n+1}{\eps}.
    \]
\end{proof}

\section{Supporting Deletions in the Low-Load Structure}
\label{app:deletions}

In this appendix, we prove \Cref{lem:low-load-deletions}, extending the low-load structure of \Cref{sec:backup-table} to support deletions. The goal is to preserve its scan bounds: $O(|x|)$ for positive queries, $O(1)$ for negative queries, and $O(|x|\log^{1+\delta}|x|)$ for insertions and deletions, with positive queries and updates analyzed in amortized expectation. As in the main text, the primary table $\HT$, membership table $\member$, and backup table $\backup$ each have $m$ cells and are interleaved in a single physical array. They are initially empty, and the total volume of insertions is at most $m/64$, including items subsequently deleted. 

\paragraph{Why Deletion Marks Alone Do Not Suffice.}
The simplest approach would be to mark an item as deleted and otherwise leave it in place. This preserves the placement machinery, but repeated deletions and reinsertions of one key would create many items with the same hash. For example, repeatedly inserting key $k$ with values of slightly increasing lengths could leave many copies near $c(k)$, even within a single size class. The analyses of \Cref{sec:erasure-oracle,sec:backup-table} rely on independence across keys and cannot treat these copies as independent items.

Instead, we keep at most one copy of each key in each size class, and reuse that copy whenever the key is reinserted into the same class.
As we describe below, the insertion algorithm will enforce this property by searching all levels of the relevant size class for an existing copy before creating a new one.
To make this possible, we also replace the exact-length priority order by an order on size classes. A copy's stored length may then increase within its class without changing its priority.

\paragraph{Copies, Size Classes, and Metadata.}
The \defn{size class} of a length $s\ge1$ is $i(s)=\lfloor \log s\rfloor$. Each \defn{copy} $y$ has a key $k(y)$ and a \defn{stored length} $|y|$, and holds an item of that key padded to length $|y|$. 
We assume that the original item length can be read from its encoding, so padding can be removed after decoding. A copy is either \defn{live} or \defn{dead}; the live copies represent exactly the current set $S$.

Each key $k$ ever inserted has one permanent membership record, created on its first insertion. Its position in $\member$ is the identifier $\ID(k)$ shared by all copies of $k$. We write $\ID(y)=\ID(k(y))$ and use the priority order
\[
    y\succ z
    \iff i(|y|)<i(|z|),\quad\text{or}\quad
    i(|y|)=i(|z|)\text{ and }\ID(y)<\ID(z).
\]
Since there is at most one copy per key and size class, this is a strict total order on copies. Identifiers and size classes never change, so neither do priorities.

Primary-cell metadata is as in \Cref{sec:erasure-oracle}: the stored length, codeword index, and owner identifier. A cell's identifier and stored length together identify its copy and determine its priority.
As in \Cref{sec:backup-construction}, the backup table stores items in chunks. Deleting a copy does not free its chunk, allowing the chunk to be reused if the copy is later resurrected.
Each occupied chunk also records its copy's identifier, stored length, and a live/dead bit, using $O(1)$ words of metadata. A membership record stores $(\phi(k),F,s)$, as in \Cref{sec:backup-reduction}, together with a live/dead bit. Here $s$ is the stored length of the most recently live copy of $k$, and $F$ is the endpoint of its most recent placement in $\HT$. Deletion does not clear cells or free chunks. Dead copies continue to participate in the placement rules, but are never reinflated.

\paragraph{Searching the Backup Table.}
Unlike the insertion-only structure, we sometimes need to search for a copy that does not exist. To search class $i$ for key $k$, search each level of that class from the home chunk of $k$ until finding the desired copy or an empty chunk. These searches proceed together in a single forward pass from $\gamma(k)$, the cell of $\backup$ in $c(k)$. Dead chunks count as occupied. Within a class, keys are distinct and chunks never move, so the usual linear-probing stopping rule remains valid. We use this search both to locate live copies and to find dead copies for reuse. We fix the buddy allocator's choices independently of the hashes; since requests depend only on the numbers of new copies in each class, the allocated subarrays do also.

\paragraph{Operations.}
\begin{itemize}
    \item \textbf{$\query(k)$:} Scan $\member$ from $c(k)$ to its first empty cell, collecting live records with fingerprint $\phi(k)$. For a candidate $(\phi(k),F,s)$ at position $p$, scan $\HT$ from $c(k)$ through $F$, collecting only symbols whose headers have identifier $p$ and stored length $s$. If at least $s$ symbols survive, decode the padded item and compare its key with $k$. Otherwise, search the backup levels of class $i(s)$ for the copy with identifier $p$. Reject the candidate if no such copy is found or its key differs from $k$. If the backup search finds the matching item, reinflate it in $\HT$, keeping its identifier and stored length, and update $F$ in its membership record. Return the matching item with its padding removed, or \textsc{Not Found} if no candidate matches. In particular, a negative query whose only matching record is dead stops after the membership scan; it never decodes a dead copy of $k$.
    \item \textbf{$\delete(k)$:} Run $\query(k)$ to find the live copy and its membership record. Mark the record as dead, then find the copy in the backup levels of its class and mark its chunk as dead. The primary table is unchanged.
    \item \textbf{$\ins(x)$}, for an item of key $k$, length $s=|x|$, and class $i=i(s)$:
    \begin{enumerate}[label=(\arabic*)]
        \item Scan $\member$ from $c(k)$ to its first empty cell. For each matching fingerprint record $(\phi(k),F,s_0)$ at position $p_0$, search the backup levels of class $i(s_0)$ for the copy with identifier $p_0$, whether live or dead. A found copy whose key is $k$ identifies the permanent record of $k$; otherwise reject the candidate. Let $p$ be the record's position, or the first empty membership cell if $k$ has no record. Set $\ID(k)=p$.
        \item Search the backup levels of class $i$ for key $k$. Any copy found is dead, since $k$ is absent.
        \item If a copy $y$ is found, \defn{resurrect} it: set $s'=\max\{s,|y|\}$, pad $x$ to length $s'$, and overwrite its chunk with the padded item, marking the chunk live. The item fits in the same chunk, since $s'<2^{i+1}$ and a class-$i$ chunk has $2^{i+2}$ cells. In $\HT$, write its fresh codeword into the first $2s'$ cells available to $y$ at or after $c(k)$, as in \Cref{sec:erasure-oracle}. Its identifier and class, and hence its priority, are unchanged. If no copy is found, set $s'=s$ and insert a new copy into $\HT$ and the newest backup level of class $i$, using identifier $p$.
        \item Write the live record $(\phi(k),F,s')$ at position $p$ in $\member$, where $F$ is the endpoint of the placement in step (3). If a record already existed, overwrite it in place.
    \end{enumerate}
\end{itemize}
For completeness, if more than one membership record matches the fingerprint, we may instead scan all chunks to identify the key and its record exactly, and then perform the requested operation. Chunk keys can be compared with $k$ as they are read, keeping one partial comparison per allocated level. This uses $O(\log^2 m)$ bookkeeping words and a constant number of full forward passes. This exceptional scan will have negligible expected cost, as in \Cref{lem:dynamic-negative}.

All scans start at $c(k)$ and proceed forward. Returning to a membership record to update it, or to $c(k)$ for a reinflation, requires only another forward pass over the same interval. Thus each operation makes $O(1)$ forward passes over one contiguous physical interval starting at $c(k)$. Since the three tables are interleaved at constant stride, it suffices to bound the lengths scanned in the constituent tables.

\begin{lemma}[Basic Invariants]
\label{lem:deletion-invariants}
At all times:
\begin{enumerate}[label=(\arabic*)]
    \item Each key has at most one copy in each size class and at most one membership record. At most one of its copies is live.
    \item Stored lengths never decrease. A copy of class $i$ has stored length in $[2^i,2^{i+1})$, and hence less than twice the length of every item inserted into that copy.
    \item The sum of the stored lengths of all copies, including dead copies, is at most the total insertion volume, and hence at most $m/64$.
    \item The copies, their stored lengths and live/dead states, and the keys with membership records are determined by the operation sequence independently of the hashes.
\end{enumerate}
\end{lemma}
\begin{proof}
An insertion creates a copy only if its class contains none of that key, and creates a membership record only if the key has never had one. Existing records are always found: they never move or disappear, and lie before the first empty cell after their canonical cell. Since only absent keys are inserted, the new or resurrected copy is the key's only live copy. A resurrection takes the maximum of the old and new lengths, both in the same size class, proving (2). Each copy's stored length is the maximum of the lengths inserted into it, and different copies receive disjoint sets of insertions, proving (3). Finally, whether an insertion creates or resurrects a copy depends only on the key and size class of earlier insertions; deletions determine which copy becomes dead. This proves (4).
\end{proof}

\begin{lemma}[Prefix Occupancy with Resurrections]
\label{lem:prefix-occupancy-deletions}
\Cref{lem:cell-monotonicity,lem:prefix-occupancy} continue to hold, with item lengths replaced by current stored lengths. After a resurrection of $y$, every primary cell naming $y$ holds a symbol of its current codeword.
\end{lemma}
\begin{proof}
The priority order is fixed, cells are never cleared, and every placement writes into the first available cells from the copy's canonical cell. Thus a cell's owner can only increase in priority, and available sets only shrink. The inductive proof of prefix occupancy applies as before. The only additional case is a resurrection increasing a copy's stored length from $s$ to $s'\ge s$. Its surviving symbols form a prefix of at most $2s$ available cells, so all are among the first $2s'$ available cells overwritten by the new placement. No symbols of the previous contents remain, and the new symbols again occupy a prefix of the available cells.
\end{proof}

\begin{lemma}[Scan Lengths in the Primary Table]
\label{lem:deletion-scan}
Let $x$ be a copy of class $i$. There is a random variable $L_x^{\max}$ bounding the primary-table scan of every query, insertion, resurrection, or reinflation of $x$ over the entire sequence, with
\[
    \E[(L_x^{\max})^2]=O(4^i)
    \qquad\text{and}\qquad
    \E[L_x^{\max}]=O(2^i).
\]
\end{lemma}
\begin{proof}
Let $s$ be the final stored length of $x$, and let $Y$ contain all copies of class at most $i$ at the end of the sequence, giving each copy $y$ weight $2|y|$. Place this weight by scanning left to right, occupying a cell whenever some copy that has arrived has unplaced weight. The occupied set depends only on the total weight hashing to each cell. Define $L_x^{\max}$ as the number of cells from $c(x)$ through the $2s$-th empty cell of this placement.

At every earlier time, every copy of higher priority than $x$ belongs to $Y$ and has at most its final weight. By prefix occupancy and the occupancy domination argument of \Cref{lem:dynamic-static-domination}, its occupied cells are contained in the occupied set of $Y$. Thus every placement of $x$ finishes by the $2s$-th empty cell of that set. A query stops no later than the endpoint of $x$'s most recent placement, so the same bound applies to queries.

Copies of one key share a hash. To recover independence, merge all copies of $Y$ with a given key into one weighted item at that key's canonical cell. By \Cref{lem:deletion-invariants}(1), its weight is less than
\[
    \sum_{j=0}^{i}2\cdot 2^{j+1}<2^{i+3}.
\]
Merging preserves the arrival weight at every cell, hence the occupied set. The merged items have distinct keys and independent hashes, and their weights are fixed independently of those hashes by \Cref{lem:deletion-invariants}(4). Their total weight is at most $m/32$.

Now add $2s$ to the merged item with key $k(x)$. The additional words fill the first $2s$ previously empty cells at or after $c(x)$, so the resulting run containing $c(x)$ has length at least $L_x^{\max}$. Every merged item still has weight $O(2^i)$, and the total weight is at most $m/32+2s\le m/16$. The run-length argument of \Cref{lem:simple-run-length}, using blocks of $O(2^i)$ cells and conditioning on $c(x)$, therefore gives a second moment of $O(4^i)$. It uses only the bounded item weights, independence of the other keys' hashes, and the fact that every word of a run hashes into that run. This bounds $\E[(L_x^{\max})^2]$; the expectation bound follows by Jensen's inequality.
\end{proof}

\begin{lemma}[Membership and Backup Scan Bounds]
\label{lem:deletion-aux}
\begin{enumerate}[label=(\arabic*)]
    \item For each key $k$, all membership scans from $c(k)$ through the first empty cell are bounded by one random variable with $O(1)$ expectation and second moment.
    \item The backup density bound of \Cref{lem:backup-density} continues to hold. For each key $k$ and class $i$, all searches of that class and all writes of a copy of $k$ in that class have scan length at most $\Lambda_{k,i}$ from $\gamma(k)$, where
    \[
        \E[\Lambda_{k,i}^2]
        =O\!\left((2^i\log^{1+\delta}2^i)^2\right).
    \]
\end{enumerate}
\end{lemma}
\begin{proof}
For (1), membership records form a unit-size linear-probing table with one permanent record per key ever inserted. There are at most $m/64$ such keys, and their set is independent of the hashes. The final table dominates every earlier membership scan, since records are never moved or removed. Applying the run-length argument of \Cref{lem:simple-run-length} with unit weights gives both moment bounds, whether or not $k$ has a record.

For (2), each chunk of class $i$ holds a copy of stored length at least $2^i$. The total stored length, including dead copies, is at most $m/64$. Thus the proof of \Cref{lem:backup-density} applies with copies in place of items. Resurrections use existing chunks and do not allocate levels.

For the scan bound, let $J_\ell$ be the number of chunks from the home chunk of $k$ through the first empty chunk in the final state of level $\ell$ of class $i$. The level contains copies with distinct keys and independent home chunks. Which copies enter the level depends only on the operation sequence: allocation and insertion into a new level depend on the number of new copies, while resurrections rewrite chunks in place. The level is at most half full. Hence the unit-weight argument of \Cref{lem:backup-cost} gives $\E[J_\ell^2]=O(1)$, including when $k$ itself has a copy in the level.

Chunks are never freed, so $J_\ell$ dominates the corresponding search length at every earlier time. A search reads at most $J_\ell$ chunks of level $\ell$, all within distance $(J_\ell+1)2^{i+2}/D_{i,\ell}$ of $\gamma(k)$. The same bound covers writing a new chunk or rewriting a resurrected one. Consequently, we may take
\[
    \Lambda_{k,i}
    =\sum_{\ell}(J_\ell+1)\frac{2^{i+2}}{D_{i,\ell}},
\]
where the sum is over the final levels of class $i$. Since $D_{i,\ell}\ge 2^\ell d_i$, Minkowski's inequality yields
\[
    \sqrt{\E[\Lambda_{k,i}^2]}
    \le \sum_{\ell}O\!\left(\frac{2^{i+2}}{2^\ell d_i}\right)
    =O(2^i/d_i)
    =O(2^i\log^{1+\delta}2^i).
\]
This proves the claimed bound uniformly over the operation sequence, including searches for absent copies.
\end{proof}

\begin{proof}[Proof of \Cref{lem:low-load-deletions}]
We first bound scan costs. An item $x$ held in a class-$i$ copy has $2^i\le |x|$, and that copy's stored length is less than $2|x|$. Thus bounds in terms of $2^i$ give the desired bounds in terms of the item length. Fingerprint collisions contribute only $O(1)$ expected scan cost per operation, by the same argument as \Cref{lem:dynamic-negative}: fingerprints are independent of the placement hash, a collision with any record has probability at most $m^{1-C}$, and even scanning the entire array costs only $O(m)$. We omit this contribution below.

\emph{Queries.} A negative query scans only $\member$, with expected scan cost $O(1)$ by \Cref{lem:deletion-aux}(1). A positive query for a class-$i$ copy $x$ scans $O(1)$ expected membership cells and at most $L_x^{\max}$ primary cells, for expected scan cost $O(2^i)$. As in \Cref{lem:backup-reduction}, we charge any backup lookup and lazy reinflation to the insertion that first created this copy.

To justify this charge across deletions and resurrections, let $K_x$ be the total number of lazy reinflations of $x$. A fallback occurs only when more than the current stored length, and hence at least $2^i$, of its symbols have been overwritten since its most recent placement. By prefix occupancy, the next placement extends its endpoint by at least $2^i$ cells. Resurrections cannot move the endpoint to the left: the available set only shrinks, and the number of symbols to be placed never decreases. Every endpoint is bounded by $L_x^{\max}$, so
\[
    K_x\le L_x^{\max}/2^i.
\]
After accounting for the initial membership scan, each fallback and reinflation contributes at most $O(1+\Lambda_{k(x),i}+L_x^{\max})$ additional scan cost. Updating the membership record accesses only the interval already scanned. By Cauchy--Schwarz and \Cref{lem:deletion-scan,lem:deletion-aux}, the total expected scan cost charged to this copy is at most
\[
\begin{aligned}
    O\!\left(\E\!\left[K_x(1+\Lambda_{k(x),i}+L_x^{\max})\right]\right)
    &\le \frac{O(1)}{2^i}
    \left(\E[L_x^{\max}]
    +\sqrt{\E[(L_x^{\max})^2]\E[\Lambda_{k(x),i}^2]}
    +\E[(L_x^{\max})^2]\right)\\
    &=O(2^i\log^{1+\delta}2^i).
\end{aligned}
\]
The insertion that created the copy had length at least $2^i$, so it pays this charge once, including all later resurrections. Dead copies are never reinflated. This leaves positive queries with amortized expected scan cost $O(|x|)$.

\emph{Insertions.} Consider an insertion of length $|x|$ in class $i$. Step (1) scans $O(1)$ expected membership cells. If the key already has a record of stored length in class $i_0$, verifying that record costs at most $\Lambda_{k,i_0}$ in the backup table. We charge this verification to the preceding insertion of $k$, whose length was at least $2^{i_0}$. That insertion receives at most one such charge, since the next insertion overwrites the record; intervening queries change only its endpoint field. Step (2) costs at most $\Lambda_{k,i}$. Step (3) scans at most $L_y^{\max}$ primary cells for the new or resurrected copy $y$, and at most $\Lambda_{k,i}$ backup cells. Step (4) updates the record within the membership interval already scanned. These bounds and the charges from lazy reinflations give amortized expected scan cost $O(|x|\log^{1+\delta}|x|)$ per insertion.

\emph{Deletions.} Deleting an item $x$ consists of a query and one additional backup search in its size class, followed by updates to the two live/dead bits. The bit updates lie in the intervals already scanned. The additional expected scan cost is $O(2^i\log^{1+\delta}2^i)$, so the amortized expected deletion cost is $O(|x|\log^{1+\delta}|x|)$.

Finally, all accesses make $O(1)$ forward passes over a contiguous interval starting at $c(k)$. Under the cache assumptions of \Cref{cor:intro-dynamic-low-em}, the scan property converts these scan bounds to the corresponding $1+O(L/B)$ I/O bounds stated in \Cref{lem:low-load-deletions}.
\end{proof}
\section{Rebuilding in Graveyard Catalan Hashing}\label{sec:graveyard-rebuild}
In this section, we prove bounds on the amortized cost of rebuilds in graveyard Catalan hashing from \Cref{sec:dynamic-high-load}. We do this by arguing that rebuilds can be accomplished locally and with $O(1)$ passes over the data structure.

\subsection{Why Rebuilding Is Local}
\label{sec:graveyard-locality}

A rebuild removes the flagged items, integrates the live items of $\CHT_{\mathrm{small}}$ into $\HT_{\mathrm{full}}$, and empties both auxiliary structures. Rebuilding by sorting all live items in external memory would be too expensive. Instead, we will rebuild independent intervals of the table inside the cache.

Let $T$ be the set of items in the old static table, including flagged items, and let $Y$ be the live set of the insertion table. For this argument, keep both copies when a key occurs in both sets, and write $U=T\uplus Y$. The new live set is $S\subseteq U$, obtained by removing the flagged items. The budget and space bounds imply
\[
    \Vol(U)\le (1-\eps/2)m+\eps'm/64
    \le (1-\eps/4)m.
\]
Let $P_U$ be any placement of $U$ that scans from left to right, placing one word whenever an arrived item still has unplaced words. The occupied cells depend only on the total volume hashing to each cell, not on which pending item is chosen. Call its maximal occupied intervals the \defn{$U$-runs}.

\begin{claim}[Rebuild Locality]
\label{claim:rebuild-locality}
For each $U$-run $R$, the old static items stored in $R$ are exactly the items of $T$ whose canonical cells lie in $R$. The new static items hashing into $R$ occupy only cells of $R$, and their layout and metadata can be computed from those items alone.
\end{claim}
\begin{proof}
Adding arrival work can only increase the occupied set. Since $T,S\subseteq U$, every run of either the old or the new static table lies within a $U$-run. An item and its canonical cell lie in the same run, proving the placement statements.

The ARCH priorities within each new run are determined by the hashes in that run. Consequently, the priorities, placement, home bits, and stop marks within $R$ depend only on the items of $S$ hashing into $R$.
\end{proof}

We next show that every such interval fits in the rebuild scratch region with high probability. Only a simple large-deviation bound is needed: a long run requires an interval to receive at least its own length in arrival work, even though the expected volume is smaller by a factor $1-\Theta(\eps)$.

\begin{lemma}[A Tail Bound for the Longest Run]
\label{lem:rebuild-run-tail}
Let $Z$ be a fixed set of items with independent uniform hashes, maximum length at most $\ell$, and volume at most $(1-\beta)m$, where $\beta\in(0,1)$. Place its words whenever arrival work is pending, as above. For every integer $r\ge1$,
\[
    \Pr[\text{some run has length at least }r]
    \le m^2\exp\!\left(-c\beta^2 r/\ell\right)
\]
for an absolute constant $c>0$.
\end{lemma}
\begin{proof}
For a fixed interval $I$, let $W(I)$ be the total length of the items with canonical cells in $I$. It is a sum of independent variables in $[0,\ell]$ with expectation at most $(1-\beta)|I|$. A Chernoff bound gives
\[
    \Pr[W(I)\ge |I|]\le\exp\!\left(-c\beta^2|I|/\ell\right).
\]
Every word of a run belongs to an item whose canonical cell lies in that run. Thus, if $I$ is a run, $W(I)=|I|$. There are at most $m^2$ intervals, so a union bound over those of length at least $r$ proves the result.
\end{proof}

\begin{claim}[All Rebuild Intervals Fit in Cache]
\label{claim:rebuild-run-length}
For a sufficiently large constant $c_0$, with probability at least $1-m^{-2}$, every $U$-run has length at most
\[
    M'=c_0s_{\max}\eps^{-2}\log m.
\]
\end{claim}
\begin{proof}
There are at most two items of $U$ with a given key: one in $T$ and one in $Y$. Merge them into a single item of length at most $2s_{\max}$. This preserves the arrival work, hence the occupied cells of $P_U$. The merged items have distinct keys and independent hashes, and their lengths and membership are fixed by the operation sequence, independently of the hashes. Apply \Cref{lem:rebuild-run-tail} with $\ell=2s_{\max}$, $\beta=\eps/4$, and $r=M'$.
\end{proof}

Only the per-item query bounds of the static table enter the operation analysis. Its expected maximum stretch is not needed.

\subsection{Rebuilding with Sequential Passes}
\label{sec:graveyard-sequential}

Locality reduces the rebuild to two tasks: enumerate the live insertion-table items in hash order, and identify the $U$-runs. Once these are available, each new static run can be computed in cache and written over the old table.

\begin{lemma}[Streaming Rebuild]
\label{lem:streaming-rebuild}
A rebuild has expected scan cost $O(m)$. Under the cache-size assumption of \Cref{sec:dynamic-high-load}, \(M\ge C_M(B+s_{\max}\eps^{-2}\log^3m),\) it takes $O(m/B)$ expected I/Os. With probability $1-O(m^{-2})$, it uses a constant number of sequential passes and $O(\eps m)$ words of temporary space outside the table; otherwise it uses $O(m)$ temporary words.
\end{lemma}
\begin{proof}
We may assume $\delta\le1$, since proving the update bound for a smaller $\delta$ suffices. First consider a long update. Its length, or the length of its deleted key, is at least $\eps'm/128$. Hence $s_{\max}\ge\eps m/(128C)$, and for sufficiently large $C_M$ the cache assumption puts the entire physical array in cache. We read it, apply the update, rebuild, and write it back in $O(m/B)$ I/Os. We now consider budget-triggered rebuilds.

\paragraph{Pass 1: Extract the Insertions in Hash Order.}
Every live item of $\CHT_{\mathrm{small}}$ has a live chunk in $\backup_{\mathrm{small}}$ (\Cref{app:deletions}). Scan the physical array, reading these chunks and ignoring the primary codewords. The allocator state identifies the level, chunk, and offset of each backup cell. A scan reads at most one partial chunk per allocated subarray at a time. There are $O(\log m)$ levels per class, and a class-$i$ chunk has $2^{i+2}$ cells. Summing over classes, the partial chunks occupy $O(s_{\max}\log m)$ cached words.

When a live chunk is complete, remove its padding, compute its hash, and keep its item in a buffer ordered by hash. Release an item $y$ to a sequential temporary stream when the scan has advanced
\[
    \lambda=C_1\sigma s_{\max}\log^{2+\delta}m
\]
physical cells past $c(y)$, for a sufficiently large constant $C_1$. The placement argument of \Cref{lem:backup-cost,lem:deletion-aux} bounds the distance to the end of a class-$i$ chunk in level $\ell$ by
\[
    O\!\left(\sigma(J_\ell+1)s_{\max}\log^{1+\delta}m\right),
\]
where $J_\ell$ has an exponential tail: the level is a unit-size linear-probing table at load at most $1/2$. A union bound over the at most $m$ live items shows that, with probability $1-O(m^{-2})$, every item is complete before its release point. When $\lambda<m_{\mathrm{total}}$, unwrap canonical positions into the interval $[m_{\mathrm{total}}-\lambda,2m_{\mathrm{total}}-\lambda)$. Use each chunk's first complete occurrence after its represented canonical position, ignoring earlier occurrences and later duplicates. The first lap initializes the cyclic buffer; the second emits one lap of releases, processing equal-cell releases in hash order. This produces a stream of $Y$ in cyclic hash order. When $\lambda\ge m_{\mathrm{total}}$, instead collect all of $Y$ in at most two laps, sort it in cache, and write the stream directly.

On this event, buffered items have canonical cells in an interval of at most $\lambda$ physical cells behind the scan. The set $Y$ has independent hashes and volume $O(\eps m)$. A Chernoff bound and a union bound over scan positions show that, with probability $1-O(m^{-2})$, the buffer uses at most
\[
    O(\eps\lambda+s_{\max}\log m)
    =O(s_{\max}\log^{2+\delta}m)
\]
words. If $\lambda$ exceeds the array length, the same bound follows by buffering all of $Y$.

\paragraph{Pass 2: Find a Boundary and Check Run Lengths.}
Start at any empty cell of the old static table and read one old run at a time. Parse each run in cache, reconstruct its items, and sort them by hash. An oversized old run triggers the fallback described below. Otherwise, merge its arrivals with the stream from pass 1 to obtain, in cyclic cell order,
\[
    a(j)=\sum_{y\in U:\,c(y)=j}|y|.
\]
If the starting cell is indexed $0$, one lap computes a minimizer $j^*$ of
\[
    F(j)=\sum_{t=0}^{j}a(t)-(j+1).
\]
Every cyclic interval ending at $j^*$ receives at most its length in work, since $F(j^*)$ is minimal and $\Vol(U)<m$. Thus no work is pending after cell $j^*$. Continuing from there with zero pending work, the recursion
\[
    w(j+1)=\max\{0,w(j)+a(j)-1\}
\]
identifies the occupied cells of $P_U$: cell $j$ is occupied exactly when $w(j)+a(j)>0$. In one additional lap, find an empty cell $e_0$ and check that every $U$-run has length at most $M'$. If a run crosses the starting point of this lap, combine its initial and final occupied segments when checking its length. All size checks take place before any table cells are overwritten.

\paragraph{Pass 3: Rebuild Each Interval in Cache.}
Starting just after $e_0$, read the table for one lap, using the same arrival-work recursion. For each $U$-run $R$, collect the old items stored in $R$ and the items of $Y$ hashing into $R$. Their total volume is $|R|\le M'$. Discard flagged items and compute the new ARCH layout in cache: priorities in decreasing cyclic hash order, delimiters, home bits, and stop marks. By \Cref{claim:rebuild-locality}, this is exactly the restriction of the new static table to $R$.

Also compute the collision bits for the new static set: initialize them to zero, sort the retained items by descriptor in cache, and mark each canonical cell whose descriptors include a duplicate. All keys with a given canonical cell lie in the same $U$-run, so this detects every relevant collision. The descriptors use $O(|R|)$ cached words and require no additional I/Os. Clear all deletion bits.

Write the new contents over the old run. The write pointer trails the read pointer, so each run is read completely before it is overwritten. Clear all cells outside the $U$-runs, both auxiliary structures, their allocator state, and the budget counters. The backup chunks have already been extracted in pass 1.

\paragraph{Space, Failure Events, and Cost.}
The cached working space is
\[
    O(s_{\max}\log^{2+\delta}m+M')
    =O(s_{\max}\eps^{-2}\log^3m),
\]
and the temporary stream contains $O(\Vol(Y))=O(\eps m)$ words. Each pass scans at most two laps of the array and a constant number of sequential streams, using $O(1)$ blocks outside the scratch region. Thus the passes cost $O(m/B)$ I/Os and scan $O(m)$ cells.

The failure events are a late chunk, an oversized extraction buffer, or an oversized run. A chunk's cyclic displacement from its canonical position is checked when it is read, so a late chunk is detected even if its chosen occurrence falls beyond the second lap. Their total probability is $O(m^{-2})$, and they are detected in the first two passes, while the old table is intact. In that case, extract the items into external working space and build the new table directly, using $O(m)$ temporary words and $O(m\log m)$ RAM operations and I/Os. This fallback contributes only $O(m^{-1}\log m)=O(m/B)$ expected I/Os for $B\le m$. For $B>m$, the whole array fits in cache under our assumptions. Charge the first access to the triggering update; once that block is cached, random rotation gives $O(m/B)$ expected additional I/Os to load the array and rebuild it there. The same estimates give expected scan cost $O(m)$, including the fallback.
\end{proof}

\end{document}